\documentclass[fleqn]{LMCS}
\usepackage{latexsym}   
\usepackage{enumerate}
\usepackage{hyperref}

\newcommand{\dlts}{L$\!^2$TS}
  \newcommand{\Acts}[1][A]{\ensuremath{\mathalpha{#1}}}
  \newcommand{\act}[1]{\ensuremath{\mathalpha{#1}}}
    \newcommand{\acta}[1][]{\ensuremath{\act{a_{#1}}}}

  \newcommand{\sact}{\act{\tau}}
  \newcommand{\silent}[1][\tau]{\ensuremath{\mathalpha{#1}}}
  \newcommand{\States}[1][S]{\ensuremath{\mathalpha{#1}}}
  \newcommand{\state}[1]{\ensuremath{\mathalpha{#1}}}
    \newcommand{\states}[1][]{\state{s_{#1}}}
    \newcommand{\statet}[1][]{\state{t_{#1}}}
    \newcommand{\stateu}[1][]{\state{u_{#1}}}
    \newcommand{\statev}[1][]{\state{v_{#1}}}

  \newcommand{\brelsym}[1][]{\ensuremath{\mathalpha{\mathcal{R}_{#1}}}}
  \newcommand{\brel}[1][]{\ensuremath{\mathrel{\brelsym[#1]}}}
   \newdimen\boxwdplusemdimen
   \def\arrow#1{{
     \boxwdplusemdimen=1em%
     \setbox0=\hbox{$\scriptstyle#1$}%
     \advance\boxwdplusemdimen by \wd0\relax%
     \ifdim\boxwdplusemdimen<16.11119pt%
       \boxwdplusemdimen=16.11119pt%
     \fi%
     \buildrel{#1}\over%
       {\setbox1=\hbox to \boxwdplusemdimen{\rightarrowfill}%
     \ht1=0.3em\relax\box1}%
   }}
   \makeatletter
   \def\twoheadrightarrowfill{$\m@th\smash-\mkern-7mu%
     \cleaders\hbox{$\mkern-2mu\smash-\mkern-2mu$}\hfill
     \mkern-7mu\mathord\twoheadrightarrow$}
   \makeatother
   \def\darrow#1{{
     \boxwdplusemdimen=1em%
     \setbox0=\hbox{$\scriptstyle#1$}%
     \advance\boxwdplusemdimen by \wd0\relax%
     \ifdim\boxwdplusemdimen<16.11119pt%
       \boxwdplusemdimen=16.11119pt%
     \fi%
     \buildrel{#1}\over%
       {\setbox1=\hbox to \boxwdplusemdimen{\twoheadrightarrowfill}%
     \ht1=0.3em\relax\box1}%
   }}
  \newcommand{\step}[1]{\ensuremath{\mathbin{\arrow{#1}}}}
  \newcommand{\stepsym}{\ensuremath{\mathalpha{\rightarrow}}}

  \newcommand{\bbisim}{%
    \setbox0=\hbox{\kern-.1ex{$\leftrightarrow$}\kern-.1ex}
    \setbox1=\vbox{\hbox{\raise .1ex \box0}\hrule}%
    \ensuremath{\mathrel{\hbox{\kern.1ex\box1\kern.1ex}_{b}}}
  }
  \newcommand{\dsbbisim}{%
    \setbox0=\hbox{\kern-.1ex{$\leftrightarrow$}\kern-.1ex}
    \setbox1=\vbox{\hbox{\raise .1ex \box0}\hrule}%
    \ensuremath{\mathrel{\hbox{\kern.1ex\box1\kern.1ex}^{\lambda}_{b}}}
  }
  \newcommand{\bbisimd}{%
    \setbox0=\hbox{\kern-.1ex{$\leftrightarrow$}\kern-.1ex}
    \setbox1=\vbox{\hbox{\raise .1ex \box0}\hrule}%
    \ensuremath{\mathrel{\hbox{\kern.1ex\box1\kern.1ex}^{\Delta}_{b}}}
  }
\newcommand{\Cc}{\bbisim}
\newcommand{\Ccl}{\dsbbisim}
\newcommand{\Ccd}{\bbisimd}
\newcommand{\dbs}{\approx_{\it dbs}}
\newcommand{\s}{\approx_s}
\newcommand{\ls}{\approx_{\lab}}
  \newcommand{\N}{\ensuremath{\mathalpha{\omega}}}

  \newcommand{\IFF}{iff}
  \newcommand{\all}[1]{\ensuremath{\mathalpha{\forall{#1}}}}
  \newcommand{\is}[1]{\ensuremath{\mathalpha{\exists{#1}}}}

  \newcommand{\compl}{\ensuremath{\mathalpha{\neg}}}
  
  \newcommand{\Meet}[1][]{\ensuremath{\bigwedge_{#1}}}
  
  \newcommand{\Until}[1][\acta]{\ensuremath{\mathbin{\mathsf{U}}}}
  
  \newcommand{\sat}{\ensuremath{\mathrel{\models}}}

\newcommand{\plat}[1]{\raisebox{0pt}[0pt][0pt]{#1}}     
\newcommand{\colours}{{\bf C}}
\newcommand{\C}{\mathcal{C}}
\newcommand{\D}{\mathcal{D}}
\newcommand{\E}{\mathcal{E}}
\newcommand{\Props}{\ensuremath{\mathbf{AP}}}
\newcommand{\aprop}[1]{\ensuremath{\mathalpha{#1}}}
  \newcommand{\propp}[1][]{\aprop{p_{#1}}}
\newcommand{\KS}[1][K]{\ensuremath{\mathalpha{#1}}}
\renewcommand{\KS}[1][K]{{\rm #1}}
\newcommand{\pth}[1]{\ensuremath{\mathalpha{#1}}}
  \newcommand{\pthp}[1][]{\ensuremath{\pth{\pi_{#1}}}}
  \newcommand{\pthr}[1][]{\ensuremath{\pth{\rho_{#1}}}}

\newcommand{\CTL}{\ensuremath{\mathsf{CTL}}}
\newcommand{\CTLminX}{\ensuremath{\mathsf{CTL}_{{-}\mathsf{X}}}}
\newcommand{\LTL}{\ensuremath{\mathsf{LTL}}}
\newcommand{\LTLminX}{\ensuremath{\mathsf{LTL}_{{-}\mathsf{X}}}}

\newcommand{\LTLD}{\ensuremath{\mathsf{LTL}_{\infty}}}
\newcommand{\CTLD}{\ensuremath{\mathsf{CTL}_{\infty}}}
\newcommand{\CTLd}{\ensuremath{\mathsf{CTL}_{\delta}}}
\newcommand{\CTLS}{\ensuremath{\mathsf{CTL}^*}}
\newcommand{\CTLSminX}{\ensuremath{\mathsf{CTL}^*_{{-}\mathsf{X}}}}
\newcommand{\CTLSD}{\ensuremath{\mathsf{CTL}^*_{\infty}}}
\newcommand{\CTLSd}{\ensuremath{\mathsf{CTL}^*_{\delta}}}
\newcommand{\SFrm}[1][\Phi]{\ensuremath{\mathalpha{#1}}}
\newcommand{\sfrm}[1]{\ensuremath{\mathalpha{#1}}}
  \newcommand{\sfrmf}[1][]{\sfrm{\varphi_{#1}}}
\newcommand{\PFrm}[1][\Psi]{\ensuremath{\mathalpha{#1}}}
\newcommand{\pfrm}[1]{\ensuremath{\mathalpha{#1}}}
  \newcommand{\pfrmp}[1][]{\pfrm{\psi_{#1}}}
\newcommand{\Pow}[1]{\ensuremath{\mathalpha{2^{#1}}}}
\newfont{\fsc}{rsfs10 scaled 1000} 
\newcommand{\Lab}{\mbox{\fsc L}}   
\newfont{\fscsmall}{rsfs10 scaled 700} 
\newcommand{\lab}{\mbox{\fscsmall 
L}}   
\newcommand{\suffsym}{\ensuremath{\unrhd}}
\newcommand{\suff}{\ensuremath{\mathrel{\suffsym}}}
\newcommand{\psuffsym}{\ensuremath{\rhd}}
\newcommand{\psuff}{\ensuremath{\mathrel{\psuffsym}}}

\newcommand{\pcmpsym}{\ensuremath{\mathalpha{\|}}}
\newcommand{\pcmp}{\ensuremath{\mathbin{\pcmpsym}}}
\newcommand{\de}{D}
\newcommand{\dec}{\mbox{\fsc{D}}}
\newcommand{\enc}{\mbox{\fsc{E}}}
\newcommand{\Id}[1]{[\hspace{-1.4pt}[#1]\hspace{-1.4pt}]} 

\usepackage[all]{xy}
\xyoption{dvips} \xyoption{color}
\newxyColor{mygray3}{0.95}{gray}{}
\newxyColor{mygray1}{0.8}{gray}{}
\newxyColor{mygray2}{0.45}{gray}{}
\newxyColor{mygray4}{0.25}{gray}{}
\newxyColor{mygray5}{0}{gray}{}
\newxyColor{mygray6}{0.65}{gray}{}

\newdir{...}{{}*{}*{}}
\newcommand{\slabl}[1]{\ar@(r,u)@{...}^{\mbox{$#1$}\hspace{.65cm}}}
\newcommand{\slabr}[1]{\ar@(l,u)@{...}_{\hspace{.65cm}\mbox{$#1$}}}
\newcommand{\newslabl}[1]{\ar@(r,u)@{...}^{\stackrel{}{\mbox{$#1$}}\hspace{.65cm}}}
\newcommand{\newslabr}[1]{\ar@(l,u)@{...}_{\hspace{.65cm}\stackrel{}{\mbox{$#1$}}}}

\newtheorem{mydefi}{Definition}
\newtheorem{myexa}{Example}
\newtheorem{mythm}{Theorem}
\newtheorem{mycor}{Corollary}
\newtheorem{mylem}{Lemma}
\newtheorem{myprop}{Proposition}
\newtheorem{myobs}{Observation}

\def\doi{5 (4:5) 2009}
\lmcsheading%
{\doi}
{1--24}
{}
{}
{Oct.~\phantom06, 2008}
{Dec.~22, 2009}
{}   

\begin{document}

\title[CTL with Deadlock Detection]%
      {Computation Tree Logic with Deadlock Detection}
\author[R.~van Glabbeek]{Rob van Glabbeek\rsuper a}
\address{{\lsuper a}%
         National ICT Australia, and
         School of Comp.\ Sc.\ and Engineering,
         University of New South Wales,
         Sydney, Australia}
\email{rvg@cs.stanford.edu}
\author[B.~Luttik]{Bas Luttik\rsuper b}
\address{{\lsuper{b,c}}%
         Dept.\ of Math.\ \& Comp.\ Sc.,
         Technische Universiteit Eindhoven,
         The Netherlands}
\email{\{s.p.luttik,n.trcka\}@tue.nl}
\author[N.~Tr\v{c}ka]{Nikola Tr\v{c}ka\rsuper c}

\setcounter{footnote}{0}

\begin{abstract}

We study the equivalence relation on states of labelled transition
systems of satisfying the same formulas in Computation Tree Logic
without the next state modality ({\CTLminX}). This relation is
obtained by De Nicola \& Vaandrager by translating labelled transition
systems to Kripke structures, while lifting the totality restriction
on the latter. They characterised it as divergence sensitive branching
bisimulation equivalence.

We find that this equivalence fails to be a congruence for
interleaving parallel composition. The reason is that the proposed
application of {\CTLminX} to non-total Kripke structures lacks the
expressiveness to cope with deadlock properties that are important in
the context of parallel composition. We propose an extension of
{\CTLminX}, or an alternative treatment of non-totality, that fills
this hiatus. The equivalence induced by our extension is characterised
as branching bisimulation equivalence with explicit divergence, which
is, moreover, shown to be the coarsest congruence contained in
divergence sensitive branching bisimulation equivalence.
\end{abstract}

\keywords{temporal logic, deadlock, parallel composition,
  stuttering equivalence, branching bisimulation equivalence,
  explicit divergence}
\subjclass{F.4.1, D.2.4}

\maketitle

\section{Introduction}\label{sect:introduction}

\noindent
{\CTLS} \cite{EH86} is a powerful state-based temporal logic combining
linear time and branching time modalities; it generalises the
branching time temporal logic {\CTL} \cite{EC82}.  {\CTLS} is
interpreted in terms of Kripke structures, directed graphs together
with a labelling function assigning to every node of the graph a set
of atomic propositions. As the \emph{next state} modality {\sf X} is
incompatible with abstraction of the notion of state, it is often
excluded in high-level specifications. By \CTLSminX{} we denote
{\CTLS} without this modality. To characterise the equivalence induced
on states of Kripke structures by validity of \CTLSminX{} formulas,
Browne, Clarke \& Grumberg \cite{BCG88} defined the notion of
\emph{stuttering equivalence}. They proved that two states in a finite
Kripke structure are stuttering equivalent if and only if they
satisfy the same \CTLSminX{} formulas, and moreover, they established
that this is already the case if and only if the two states satisfy
the same {\CTLminX} formulas.

There is an intuitive correspondence between the notions of stuttering
equivalence on Kripke structures and \emph{branching bisimulation
equivalence} \cite{GW96} on labelled transition systems (LTSs),
directed graphs of which the edges are labelled with actions. De
Nicola \& Vaandrager \cite{DV95} have provided a framework
for constructing natural translations between LTSs and Kripke structures
in which this correspondence can be formalised. Stuttering equivalence
corresponds in their framework to a \emph{divergence sensitive}
variant of branching bisimulation equivalence, and conversely,
branching bisimulation equivalence corresponds to a \emph{divergence
  blind} variant of stuttering equivalence. The latter characterises
the equivalence induced on states of Kripke structures by a divergence
blind variant of validity of \CTLSminX{} formulas.

In \cite{EC82,EH86,BCG88} and other work on \CTLS{}, Kripke structures are
required to be \emph{total}, meaning that every state has an outgoing
transition. These correspond with LTSs that are \emph{deadlock-free}.
In the world of LTSs requiring deadlock-freeness is considered a
serious limitation, as deadlock is introduced by useful process
algebraic operators like the \emph{restriction} of CCS and the
\emph{synchronous parallel composition} of CSP\@. Conceptually, a
deadlock may arise as the result of an unsuccessful synchronisation
attempt between parallel components, and often one wants to verify
that the result of a parallel composition is deadlock-free.  This is,
of course, only possible when working in a model of concurrency
where deadlocks can be expressed.

Through the translations of \cite{DV95} it is possible to define the
validity of {\CTLSminX} formulas on states of LTSs\@.  To apply
{\CTLSminX}-formulas to LTSs that may contain deadlocks, De Nicola \&
Vaandrager \cite{DV95} consider Kripke structures with deadlocks as
well, and hence lift the requirement of totality.  They do so by using
maximal paths instead of infinite paths in the definition of validity
of {\CTLSminX} formulas.  Without further changes, this amounts to the
addition of a self-loop to every deadlock state.  As a consequence,
{\CTLSminX} formulas cannot see the difference between a state without
outgoing transitions (a \emph{deadlock}) and one whose only outgoing
transition constitutes a self-loop (a \emph{livelock}), and
accordingly a deadlock state is stuttering equivalent to a livelock
state that satisfies the same atomic propositions. This paper will
challenge the wisdom of this set-up.

We observe that for systems with deadlock, the divergence sensitive
branching bisimulation equivalence of \cite{DV95} fails to be a
congruence for interleaving operators. We characterise the coarsest
congruence contained in divergence sensitive branching bisimulation
equivalence as the \emph{branching bisimulation equivalence with
explicit divergence} introduced in \cite{GW96}. This equivalence
differs from divergence sensitive branching bisimulation equivalence
in that it distinguishes deadlock and livelock.  For deadlock-free
systems the equivalences coincide.

Having established that the framework of \cite{DV95} turns {\CTLSminX}
into a logic on LTSs that induces an equivalence under which
interleaving parallel composition fails to be compositional, we
propose two adaptations to this framework that both make {\CTLSminX}
induce branching bisimulation equivalence with explicit divergence and
thus restore compositionality.  Our first adaptation preserves the
treatment of non-totality of \cite{DV95} as well as their translations
between LTSs and Kripke structures, but extends the language
{\CTLSminX} so that it can distinguish deadlock from successful
termination. Our second adaptation preserves the totality requirement
on Kripke structures but modifies the translation from LTSs to Kripke
structures. One of our main results is that both adaptations are
equivalent in the sense that they induce equally expressive logics on
LTSs. In the following two paragraphs we discuss these adaptations in
more detail.

That divergence sensitive branching bisimulation equivalence is not a
congruence for interleaving operators means that there are properties
of concurrent systems, pertaining to their deadlock behaviour, that
(in the framework of \cite{DV95})
cannot be expressed in {\CTLSminX}, but that can be expressed in terms
of the validity of a {\CTLSminX} formula on the result of putting these
systems in a given context involving an interleaving operator.
We find this unsatisfactory, and therefore propose an extension of
{\CTLSminX} in which this type of property can be expressed
directly. We obtain that two states are branching bisimulation
equivalent with explicit divergence if and only if they satisfy the
same formulas in the resulting logic.
Treating {\CTLminX} in the same way leads either to an extension of
{\CTLminX} or, equivalently, to a modification of its semantics.  The
new semantics we propose for {\CTLminX} is a valid extension of the
original semantics \cite{EC82} to non-total Kripke structures. It
slightly differs from the semantics of \cite{DV95} and it is arguably
better suited to deal with deadlock behaviour.

Instead of extending {\CTLSminX} or modifying {\CTLminX} we also
achieve the same effect by amending the translation from LTSs to
Kripke structures in such a way that every LTS maps to a total Kripke
structure.  This amended translation consist of any of the
translations in the framework of \cite{DV95} followed by a
postprocessing stage introducing a fresh state $s_\delta$, labelled by
a fresh atomic proposition expressing the property of having
deadlocked, and a transition from all deadlock states, and $s_\delta$
itself, to $s_\delta$.  Adding self-loops and a fresh atomic
proposition expressing deadlock (or just a fresh atomic proposition
expressing deadlock) to deadlock states themselves does not have the
desired effect, for it yields logics that are too expressive.

From the point of view of practical applications our work allows the
rich tradition of verification by equivalence checking to be combined
with the full expressive power of \CTLSminX. In equivalence checking,
three properties of the chosen equivalence have been found
indispensable \cite{handbook}: compositionality---in particular
parallel composition being a congruence---is a crucial requirement to
combat the state explosion problem; the ability to represent deadlock
is crucial in ascertaining deadlock-freedom; and abstraction from
internal activity---and thus from the concept of a ``next state''---is
crucial to get a firm grasp of correctness. Our work is the first that
allows specification by arbitrary \CTLSminX{} formulas to be
incorporated in this framework, without giving up any of these
essential properties.

Given the existence of adequate translations between LTSs and Kripke
structures, we could have presented the results of this paper entirely
within the framework of Kripke structures, or entirely within the
framework of LTSs. Using Kripke structures only would entail defining
a parallel composition on Kripke structures---which is possible by
lifting the parallel composition on LTSs through the appropriate
translations. However, Kripke structures are traditionally used for
global descriptions of systems; building system descriptions modularly
by parallel composition, while worrying about deadlocks that may be
introduced in this process, would be a novel approach in itself. For
establishing the results of this paper it is much more appropriate to
build on the rich tradition of composing LTSs by parallel composition,
and the known importance of deadlock behaviour within this framework.

Using just LTSs, on the other hand, would require lifting \CTLSminX{}
to the world of LTSs.\footnote{A tempting alternative appears to
be to use the \emph{weak modal $\mu$-calculus} \cite{RS97} instead of
\CTLSminX{}. This is the modal $\mu$-calculus of Kozen \cite{Koz83} with
\emph{weak} action modalities $\langle\!\langle \alpha \rangle\!\rangle$
and $[\hspace{-1.25pt}[a]\hspace{-1.25pt}]$ instead of $\langle a \rangle$
and $[a]$ in order to abstract from internal activity. However, as
observed in \cite{RS97}, this logic cannot distinguish states that
are \emph{weakly bisimilar}, and hence, contrary to what is suggested
in  the introduction of \cite{RS97}, lacks the expressiveness of \CTLSminX.}
Here we could build on the work of De Nicola and
Vaandrager \cite{DV90}, who defined the logic $\mathsf{ACTL}^*$ on
LTSs and showed that it corresponds neatly, through the translations
of \cite{DV95}, with $\CTLS$ on Kripke structures. However, whereas
abstracting from the notion of state in {\CTLS} can be done elegantly
by removing the next state modality {\sf X} from the language, in
$\mathsf{ACTL}^*$ this additionally requires parametrising the
\emph{until}-modality by two \emph{action formulas} \cite{DV90}. Doing
this would make the resulting logic $\mathsf{ACTL}^*_{-\mathsf{X}}$
appear less than a wholly canonical action-based incarnation of
$\CTLSminX$, and the reader might wonder whether the failure of
$\mathsf{ACTL}^*_{-\mathsf{X}}$ to generate an equivalence on LTSs
that is a congruence for parallel composition would be due to it being
an imperfect rendering of $\CTLSminX$ in the action-based world.

By presenting our analysis directly for $\CTLSminX$, we make clear that this is
not the case, and the problem stems from $\CTLSminX$ itself.  Having to work in
both LTSs and Kripke structures, with translations between them, appears to be
a small price to pay. In addition, we feel that in many applications, such as
process algebra with data, in may be preferable to work directly in a model of
concurrency that features both state and action labels, and thus benefits from
the ability to smoothly combine LTSs and Kripke structures~\cite{TrckaPhD}.

Nevertheless, all our work applies just as well to
$\mathsf{ACTL}^*_{-\mathsf{X}}$, with the very same problems and the
very same solutions.

At the end of the paper we briefly consider Linear Temporal Logic
without the next state modality (\LTLminX{}). The equivalence induced
by the validity of \LTLminX{}-formulas is not a congruence for
interleaving parallel composition either. The coarsest coarsest
congruence included in the equivalence induced by the validity of
\LTLminX{}-formulas is obtained much in the same way as the coarsest
congruence included in the equivalence induced by the validity of
\CTLminX{}-formulas. Adding the $\infty$-modality to \LTLminX{},
however, yields a logic that induces a strictly finer equivalence than
the obtained congruence.

\section[]{\texorpdfstring{\CTLSminX{}}{CTL*X} and stuttering equivalence}
\label{CTL}

\noindent
We presuppose a set $\Props$ of \emph{atomic propositions}.
A \emph{Kripke structure} is a tuple
$(\States,\Lab,\stepsym)$ consisting of
  a set of states $\States$,
  a \emph{labelling function}
    $\Lab:\States\rightarrow\Pow{\Props}$
and
  a \emph{transition relation}
    $\stepsym\subseteq\States\times\States$.
For the remainder of the section we fix a Kripke structure
  $(\States,\Lab,\stepsym)$.

A \emph{finite path} from $\states$ is a finite sequence of states
$\states[0],\dots,\states[n]$ such that $\states=\states[0]$
and $\states[k]\step{}\states[k+1]$ for all $0\leq k < n$.
An \emph{infinite path} from $\states$ is an infinite sequence of
states $\states[0],\states[1],\states[2],\dots$ such that
$\states=\states[0]$ and $\states[k]\step{}\states[k+1]$ for all
$k\in\N$.
A \emph{path} is a finite or infinite path.
A \emph{maximal path} is an infinite path
or a finite path $\states[0],\dots,\states[n]$
such that $\neg\exists\state{s'}.\ \states_n\step{}\state{s'}$.
We write $\pthp\suff\pth{\pi'}$ if the path $\pth{\pi'}$ is a suffix
of the path $\pthp$, and $\pthp\psuff\pth{\pi'}$ if
$\pthp\suff\pth{\pi'}$ and $\pthp\not=\pth{\pi'}$.

Temporal properties of states in $S$ are defined using \CTLSminX{} formulas.
\begin{defi}\rm
  The classes $\SFrm$ of \emph{\CTLSminX{} state formulas} and
  $\PFrm$ of \emph{\CTLSminX{} path formulas} are generated by the
  following grammar:
\[
    \sfrmf ::=\
      \propp\ \mid\
      \compl\sfrmf\ \mid\
      \Meet\SFrm[\Phi']\ \mid\
      \is\pfrmp
\qquad\qquad
    \pfrmp ::=\
      \sfrmf\ \mid\
      \compl\pfrmp\ \mid\
      \Meet\PFrm[\Psi'] \mid\
      \pfrmp\Until\pfrmp
\]
  with $\propp\in\Props$, $\sfrmf\in\SFrm$,
  $\SFrm[\Phi']\subseteq\SFrm$,
  $\pfrmp\in\PFrm$ and $\PFrm[\Psi']\subseteq\PFrm$.
\end{defi}
\noindent
In case the cardinality of the set of states of our Kripke
structure is less than some infinite cardinal $\kappa$,\footnote{In
  fact it suffices to require that for every state $s$ the cardinality of
  the set of states reachable from $s$ is less than $\kappa$.} we may require
that $|\SFrm[\Phi']|<\kappa$ and $|\PFrm[\Psi']|<\kappa$ in
conjunctions, thus obtaining a \emph{set} of formulas rather than a
proper class. Normally, $S$ is required to be finite, and accordingly
\CTLSminX{} admits finite conjunctions only.

\begin{defi}\rm\label{validity}
  We define when a \CTLSminX{} state formula $\sfrmf$ is \emph{valid}
  in a state $\states$ (notation: $\states\sat\sfrmf$) and when a
  \CTLSminX{} path formula $\pfrmp$ is \emph{valid} on a maximal path
  $\pthp$ (notation: $\pthp\sat\pfrmp$) by simultaneous induction as
  follows:
\begin{list}{$-$}{\leftmargin 18pt
                  \itemsep 0pt \parsep 0pt
                        \labelwidth\leftmargini\advance\labelwidth-\labelsep}
  \item $\states\sat\propp$ \IFF{} $\propp\in\Lab(\states)$;
  \item $\states\sat\compl\sfrmf$ \IFF{} $\states\not\sat\sfrmf$;
  \item $\states\sat\Meet\SFrm[\Phi']$ \IFF{}
    $\states\sat\sfrmf$ for all $\sfrmf\in\SFrm[\Phi']$;
  \item $\states\sat\is\pfrmp$ \IFF{} there exists a maximal path
    $\pthp$ from $\states$ such that $\pthp\sat\pfrmp$;
  \item $\pthp\sat\sfrmf$ \IFF{} $\states$ is the first state of
    $\pthp$ and $\states\sat\sfrmf$;
  \item $\pthp\sat\compl\pfrmp$ \IFF{} $\pthp\not\sat\pfrmp$;
  \item $\pthp\sat\Meet\PFrm[\Psi']$ \IFF{}
    $\pthp\sat\pfrmp$ for all $\pfrmp\in\PFrm[\Psi']$; and
  \item $\pthp\sat\pfrm{\psi}\Until\pfrm{\psi'}$ \IFF{}
    there exists a suffix $\pth{\pi'}$ of $\pthp$ such that
      $\pth{\pi'}\sat\pfrm{\psi'}$,
    and
      $\pth{\pi''}\sat\pfrm{\psi}$
        for all $\pthp\suff\pth{\pi''}\psuff\pth{\pi'}$.
  \end{list}
A formula $\psi\Until\psi'$ says that, along a given path, $\psi$ holds
\emph{until} $\psi'$ holds.
One writes $\top$ for the empty conjunction (which is always valid),
${\sf F}\psi$ for $\top\Until\psi$ (``$\psi$ will hold \emph{eventually}'')
and ${\sf G}\psi$ for $\neg{\sf F}\neg\psi$ (``$\psi$ holds \emph{always}
(along a path)'').
\end{defi}

\noindent
The above is the standard interpretation of \CTLSminX{}
\cite{EH86,BCG88}, but extended to Kripke structures that are not
required to be total. Following \cite{DV95}, this is achieved by using
maximal paths in the definition of validity of {\CTLSminX} formulas,
instead of the traditional use of infinite paths \cite{EH86,BCG88}.
The resulting definition generalises the traditional one, because for
total Kripke structures a path is maximal iff it is infinite.

An equivalent way of thinking of this generalisation of {\CTLSminX} to
non-total Kripke structures is by means of a transformation that makes
a Kripke structure $\KS$ total by the addition of a self-loop $s\step{}s$
to every deadlock state $s$, together with the convention that a
formula is valid in a state of $\KS$ iff it is valid in the same state of
the total Kripke structure obtained by this transformation. It is not
hard to check that this yields the same notion of validity as our
Definition~\ref{validity}.

The \emph{divergence blind} interpretation of \cite{DV95} (notation:
$\states\sat_{\it db}\sfrmf$ and $\pthp\sat_{\it db}\pfrmp$) is
obtained by dropping the word ``maximal'' in the fourth clause of
Definition~\ref{validity}.  In contrast, we call the the standard
interpretation \emph{divergence sensitive}, because it does not
abstract from \emph{divergences}, i.e., infinite paths consisting of
states with the same label. For instance, in
Figure~\ref{fig::stt-example}a we have $\statet\sat\exists {\sf G}p$,
due to the divergence $\statet,\statet,\statet,\dots$, whereas
$\stateu\not\sat\exists {\sf G}p$.  Under the divergence blind
interpretation there is no formula distinguishing these two states.
\begin{figure}\centering
$
    \begin{array}{@{}ccccccccccccccccc}
    a)
    \xymatrix@R=.9cm@C=0.3cm{
            & *++[o][F][F*:mygray1]{s} \slabl{p} \ar@/_/[ld]_{}\ar@/^/[rd]_{} & \\
          *++[o][F][F*:mygray1]{t} \slabr{p} \ar@(ld,lu)[]^{} \ar[d]_{} & & *++[o][F][F*:mygray1]{u} \slabl{p} \ar[d]_{}\\
          *++[o][F][F*:mygray2]{x} \slabr{q} & & *++[o][F][F*:mygray2]{y} \slabl{q} \ar@(rd,ru)[]^{}
          }
    & &
    b)
    \xymatrix@R=.9cm@C=0.3cm{
            & *++[o][F][F*:mygray3]{s} \slabl{p} \ar@/_/[ld]_{}\ar@/^/[rd]_{} & \\
          *++[o][F][F*:mygray1]{t} \slabr{p} \ar@(ld,lu)[]^{} \ar[d]_{} & & *++[o][F][F*:mygray4]{u} \slabl{p} \ar[d]_{}\\
          *++[o][F][F*:mygray2]{x} \slabr{q} & & *++[o][F][F*:mygray2]{y} \slabl{q} \ar@(rd,ru)[]^{}
          }
     \end{array}$
\caption{Difference between a) $\dbs$ and b) $\s$.}
 \label{fig::stt-example}
\end{figure}

\begin{defi}\rm\label{stuttering}
A \emph{colouring} is a function $\C:S\rightarrow \colours$, for
$\colours$ any set of \emph{colours}.

Given a colouring $\C$ and a (finite or infinite) path
  $\pi = s_0,s_1, s_2,\dots$ from $s$,
let $\C(\pi)$ be the sequence of colours obtained from
$\C(s_0),\C(s_1),\C(s_2),\dots$ by contracting all its (finite or
infinite) maximal consecutive subsequences $C,C,C,\dots$ to $C$. The sequence
$\C(\pi)$ is called a \emph{$\C$-coloured trace} of $s$; it is
\emph{complete} if $\pi$ is maximal.

A colouring $\C$ is \emph{[fully] consistent} if two states of the same colour
always satisfy the same atomic propositions and have the same [complete]
$\C$-coloured traces. Two states $s$ and $t$ are \emph{divergence blind
stuttering equivalent}, notation $s\dbs t$, if there exists a consistent
colouring $\C$ such that $\C(s)=\C(t)$. They are \emph{(divergence sensitive)
stuttering equivalent}, notation $s\s t$, if there exists a fully consistent
colouring $\C$ such that $\C(s)=\C(t)$. The difference between $\dbs$ and $\s$
is illustrated in the following example.
\end{defi}

\begin{exa}\rm
Consider the Kripke structure and its colouring depicted in
Figure~\ref{fig::stt-example}a. This colouring is consistent, implying $s\dbs
t\dbs u$ and $x\dbs y$, but it is not fully consistent because state $t$ has a
complete trace $\xymatrix@R=1cm@C=0.3cm{*++[o][F*:mygray1]{}}$ while $u$ does
not. Note that $t$ has, due to the self-loop, a complete coloured trace
     that consists of just the colour of a $p$-labelled state, whereas the
     unique complete coloured trace of $u$ contains the colour of a
     $q$-labelled state too. Since a consistent colouring assigns different
     colours to states with different labels, every fully consistent
     colouring must assign different colours to states $t$ and $u$, i.e.\
     it must be that $t\not\s u$. One such colouring is given in
Figure~\ref{fig::stt-example}b. This colouring shows that $x\s y$.
\end{exa}

\begin{lem}\label{2}
Let $\C$ be a colouring such that two states with the same colour
satisfy the same atomic propositions and have
the same $\C$-coloured traces of length two. Then $\C$ is consistent.
\end{lem}
\begin{proof}
Suppose $\C(s_0)=\C(t_0)$ and $C_0, C_1, C_2, \dots$ is an infinite
coloured trace of $s_0$. Then, for $i>0$, there are states $s_i$ and
finite paths $\pi_i$ from $s_{i-1}$ to $s_{i}$, such that
$\C(\pi_i)=C_{i-1},C_{i}$. With induction on $i>0$ we find
states $t_i$ with $C(s_i)=C(t_i)$ and finite paths $\rho_i$ from
$t_{i-1}$ to $t_{i}$ such that $\C(\rho_i)=C_{i-1},C_{i}$.
Namely, the assumption about $\C$ allows us to find $\rho_i$ given
$t_{i-1}$, and then $t_i$ is defined as the last state of $\rho_i$.
Concatenating all the paths $\rho_i$ yields an infinite path $\rho$
from $t_0$ with $\C(\rho)=C_0, C_1, C_2, \dots$.

The case that $\C(s_0)=\C(t_0)$ and $C_0,\dots, C_n$ is a finite
coloured trace of $s_0$ goes likewise.
\end{proof}

\begin{lem}\label{1}
Let $\C$ be a colouring such that two states with the same colour
satisfy the same atomic propositions and have
the same $\C$-coloured traces of length two, and the same complete
$\C$-coloured traces of length one. Then $\C$ is fully consistent.
\end{lem}
\begin{proof}
Suppose $\C(s)=\C(t)$ and $\sigma$ is a complete $\C$-coloured trace
of $s$. Then $\sigma=\C(\pi)$ for a maximal path $\pi$ from $s$.  By
Lemma~\ref{2}, $\sigma$ is also a $\C$-coloured trace of $t$.  It
remains to show that it is a \emph{complete} $\C$-coloured trace of
$t$.  Let $\rho$ be a path from $t$ with $\C(\rho)=\sigma$. If $\rho$
is infinite, we are done. Otherwise, let $t'$ be the last state of
$\rho$.  Then $\C(t')$ is the last colour of $\sigma$. Therefore,
there is a state $s'$ on $\pi$ such that the suffix $\pi'$ of $\pi$
starting from $s'$ is a maximal path with $\C(\pi')=\C(s')=\C(t')$.
By the assumption about $\C$, $\C(t')$ must also be a complete
$\C$-coloured trace of $t'$, i.e.\ there is a maximal path $\rho'$ from
$t'$ with $\C(\rho')=\C(t')$.  Concatenating $\rho$ and $\rho'$ yields
a maximal path $\rho''$ from $t$ with $\C(\rho'')=\sigma$.
\end{proof}

\noindent
The following two theorems were proved in \cite{DV95} and
\cite{BCG88}, respectively, for states $s$ and $t$ in a finite Kripke
structure.  Here we drop the finiteness restriction.

\begin{thm}\label{dbs char}
$s \dbs t$ iff $s\!\models_{\it db}\! \sfrmf\Leftrightarrow
t\!\models_{\it db}\!\sfrmf$ for all \CTLSminX{} state formulas $\sfrmf$.
\end{thm}
\begin{proof}
``Only if'': Let $\C$ be a consistent colouring.
With structural induction on $\sfrmf$ and $\pfrmp$
we show that
\[
  \C(s)=\C(t) \Rightarrow (s\!\models_{\it db}\! \sfrmf
\Leftrightarrow t\!\models_{\it db}\!\sfrmf) \quad\mbox{and}\quad
\C(\pi)=\C(\rho) \Rightarrow (\pi\!\models_{\it db}\! \pfrmp
\Leftrightarrow \rho\!\models_{\it db}\!\pfrmp).\vspace{-1ex}
\]
The case $\sfrmf=p$ for $p\in\Props$ follows immediately from
Definition~\ref{stuttering}. The cases
$\sfrmf=\compl\sfrmf'$ and $\sfrmf=\Meet\SFrm[\Phi']$ follow
immediately from the induction hypothesis.

Suppose $\C(s)=\C(t)$ and $s\!\models_{\it db}\! \exists \pfrmp$. Then
there exists a path $\pi$ from $s$ such that $\pi\!\models_{\it db}\!
\psi$. $\C(\pi)$ is a coloured trace of $s$, and hence of $t$. Thus
there must be a path $\rho$ from $t$ with $\C(\pi)=\C(\rho)$. By
induction, $\rho\!\models_{\it db}\!\psi$.  Hence, $t\!\models_{\it db}\!
\exists\pfrmp$.

The case $\pfrmp\in \Phi$ follows since the first states of two paths
with the same colour also have the same colour.
The cases $\pfrmp=\compl\pfrmp'$ and $\pfrmp=\Meet\SFrm[\Psi']$ follow
immediately from the induction hypothesis.

Finally, suppose $\C(\pthp)=\C(\pthr)$ and
$\pthp\models_{\it db}\pfrm{\psi}\Until\pfrm{\psi'}$.
Then there exists a suffix $\pth{\pi'}$ of $\pthp$ such that
  $\pth{\pi'}\sat_{\it db}\pfrm{\psi'}$
and
  $\pth{\pi''}\sat_{\it db}\pfrm{\psi}$
    for all $\pthp\suff\pth{\pi''}\psuff\pth{\pi'}$.
As $\C(\pthp)=\C(\pthr)$, there must be a suffix $\pth{\rho'}$ of
$\pthr$ such that $\C(\pth{\pi'}) = \C(\pth{\rho'})$ and
for every path $\pth{\rho''}$ such that
$\pthr\suff\pth{\rho''}\psuff\pth{\rho'}$ there exists a path
$\pth{\pi''}$ with $\pthp\suff\pth{\pi''}\psuff\pth{\pi'}$
such that $\C(\pi'')=\C(\rho'')$.
By induction, this implies $\pth{\rho'}\sat_{\it db}\pfrm{\psi'}$ and
$\pth{\rho''}\sat_{\it db}\pfrm{\psi}$ for all
$\pthr\suff\pth{\rho''}\psuff\pth{\rho'}$.
Hence $\pthr\models_{\it db}\pfrm{\psi}\Until\pfrm{\psi'}$.

``If'': Let $\C$ be the colouring given by $\C(s)=\{\sfrmf\mathbin\in\SFrm
\mid s\sat_{\it db}\sfrmf\}$.  It suffices to show that $\C$ is consistent.
So suppose $\C(s)=\C(t)$. Trivially, $s$ and $t$ satisfy the same
atomic propositions. By Lemma~\ref{2} it remains to show that $s$ and
$t$ have the same coloured traces of length two. Suppose $s$ has a
coloured trace $C,D$. Let
$s_0,\dots,s_k$ be a path from $s$ such that $\C(s_i)=C$ for
$0\leq i < k$ and $\C(s_k)=D\neq C$. Let\begin{center}
$\mathcal{U}=\{u\mid$ there is a path from $t$ to $u$ and $\C(u)\neq
C\}$,\\ $\mathcal{V}=\{v\mid$ there is a path from $t$ to $v$
and $\C(v)\neq D\}$.
\end{center}
For every $u\in \mathcal{U}$ pick a \CTLSminX{} formula $\sfrmf_u \in
C-\C(u)$ (using negation on a formula in $\C(u)-C$ if needed),
and for every $v\in \mathcal{V}$ pick a \CTLSminX{} formula $\sfrmf'_v
\in D-\C(v)$.  Now $s \models_{\it db} \exists (\bigwedge_{u\in
\mathcal{U}} \sfrmf_u ) \Until (\bigwedge_{v\in \mathcal{V}} \sfrmf'_v
)$ and, as $\C(s)=\C(t)$, also $t \models_{\it db} \exists (\bigwedge_{u\in
\mathcal{U}} \sfrmf_u ) \Until (\bigwedge_{v\in \mathcal{V}}
\sfrmf'_v)$.  Thus, there is a path $t_0,\dots,t_\ell$ from $t$
such that $t_\ell \models_{\it db} \bigwedge_{v\in \mathcal{V}} \sfrmf'_v$ and
$t_j \models_{\it db} \bigwedge_{u\in \mathcal{U}} \sfrmf_u$ for $0\leq
j<\ell$. It follows that $t_\ell \not\in \mathcal{V}$ and $t_j \not\in
\mathcal{U}$ for $0\leq j<\ell$. Hence $\C(t_\ell)=D$ and
$\C(t_j)=C$ for $0\leq j<\ell$, so $C,D$ is also a coloured trace of $t$.
\end{proof}

\begin{thm}\label{stuttering char}
$s \s t$ iff $s\models \sfrmf \Leftrightarrow t\models\sfrmf$
for all \CTLSminX{} state formulas $\sfrmf$.
\end{thm}
\begin{proof}
``Only if'' goes exactly as in the previous proof, reading $\sat$ for
$\sat_{\it db}$, but requiring $\C$ to be \emph{fully} consistent and,
in the second paragraph, the paths $\pi$ and $\rho$ to be maximal and
$\C(\pi)$ to be a \emph{complete} coloured trace of $s$ and $t$.

``If'' goes as in the previous proof, but this time we have
to show that $\C$ is \emph{fully} consistent. Thus, applying
Lemma~\ref{1}, and assuming $\C(s)=\C(t)$, we additionally have to
show that $s$ and $t$ have the same complete coloured traces of length
one. Let $\pi$ be a maximal path from $s$ with $\C(\pi)=C$. Let
\begin{center}
$\mathcal{U}=\{u\mid$ there is a path from $t$ to $u$ and $\C(u)\neq C\}$.
\end{center}
For every $u\!\in \mathcal{U}$ pick a \CTLSminX{} formula $\sfrmf_u
\in C-\C(u)$.  Now $s \models \exists {\sf G}(\bigwedge_{u\in
\mathcal{U}} \sfrmf_u)$ and, as $\C(s)=\C(t)$, also $t \models \exists
{\sf G} (\bigwedge_{u\in\mathcal{U}} \sfrmf_u)$.  Thus, there is a
maximal path $\rho$ from $t$ such that $t' \models \bigwedge_{u\in
\mathcal{U}} \sfrmf_u$ for all states $t'$ in $\rho$. It follows that
$t' \not\in \mathcal{U}$. Hence $\C(t')=C$ and thus $\C(\rho)=C$.
\end{proof}

\noindent
Since $\Leftrightarrow$ is an equivalence relation on predicates, we
obtain the following corollary to Theorems~\ref{dbs char} and
\ref{stuttering char}.
\begin{cor}
$\dbs$ and $\s$ are equivalence relations.\qed
\end{cor}

\noindent
Note that, for any colouring $\C$, the $\C$-coloured traces of a state
$s$ are completely determined by the complete $\C$-coloured traces of
$s$, namely as their prefixes.
Hence, any colouring that is fully consistent is certainly consistent,
and thus $\s$ is a finer (i.e.\ smaller, more discriminating)
equivalence relation than $\dbs$.

Above, the divergence blind interpretation of \CTLSminX{} is defined
by using paths instead of maximal paths. It can equivalently be
defined in terms of a transformation on Kripke structures, namely the
addition of a self-loop $s \step{} s$ for every state $s$.\footnote{
In the beginning of this section we proposed a transformation that
adds a self-loop $s \step{} s$ merely to every \emph{deadlock} state $s$.
Both transformations make any Kripke structure total. However, whereas
the previous transformation preserves the divergence sensitive
interpretation of \CTLSminX, the current one preserves the divergence
blind interpretation, and expresses it in terms of the divergence
sensitive one.}
Now $s \dbs t$ holds in a certain Kripke structure iff $s\s t$ holds in the
Kripke structure obtained by adding all these self-loops. This is
because the colour of a path doesn't change when self-loops are added
to it, and up to self-loops any path is maximal. Likewise, $s
\models_{\it db} \sfrmf$ in the original Kripke structure iff
$s\sat\sfrmf$ in the modified one.

Just like $\dbs$ can be expressed in terms of $\s$ by means of a
transformation on Kripke structures, by means of a different
transformation, at least for finite Kripke structures, $\s$ can be
expressed in terms of $\dbs$.  This is done in \cite{DV95},
Definitions~3.2.6 and~3.2.7.

\section{Branching bisimulation equivalence in terms of coloured traces}
\label{branching}

\noindent
We presuppose a set $\Acts$ of \emph{actions} with a special element
$\sact\mathbin\in\Acts$. A \emph{labelled transition system} (LTS) is a
structure $(\States,\stepsym)$ consisting of
  a set of states $\States$
and
  a \emph{transition relation}
    $\stepsym\subseteq\States\times\Acts\times\States$.
For the remainder of the section we fix an LTS $(\States,\stepsym)$.
We write $\states\step{\acta}\state{s'}$ for
$(\states,\acta,\state{s'})\in\stepsym$.

A \emph{path} from $\states$ is an alternating sequence
$\states[0],a_1,\states[1],a_2,\dots$ of states and actions, ending
with a state if the sequence is finite, such that $\states=\states[0]$
and $\states[k-1]\step{a_k}\states[k]$ for all relevant $k>0$.
A \emph{maximal path} is an infinite path
or a finite path $\states[0],a_1,\states[1],a_2,\dots,a_n,\states[n]$
such that $\neg\exists a,\state{s'}.\ \states_n\step{a}\state{s'}$.
We write $\pthp\suff\pth{\pi'}$ if the path $\pth{\pi'}$ is a suffix
of the path $\pthp$, and $\pthp\psuff\pth{\pi'}$ if
$\pthp\suff\pth{\pi'}$ and $\pthp\not=\pth{\pi'}$.

\begin{defi}\rm\label{colourings}
A \emph{colouring} is a function $\C:S\rightarrow \colours$, for
$\colours$ any set of \emph{colours}.

For $\pi = s_0,a_1,s_1,a_2,\dots$ a path from $s$, let $\C(\pi)$
be the alternating sequence of colours and actions obtained from
$\C(s_0),a_1,\C(s_1),a_2,\dots$ by contracting all finite
maximal consecutive subsequences $C,\tau,C,\tau,\dots,\tau,C$ and all
infinite maximal consecutive subsequences $C,\tau,C,\tau,\dots$ to
$C$. The sequence $\C(\pi)$ is called a
\emph{$\C$-coloured trace} of $s$; it is \emph{complete} if $\pi$ is maximal;
it is \emph{divergent} if it is finite whilst $\pi$ is infinite.

A colouring $\C$ is \emph{[fully] consistent} if two states of the
same colour always have the same [complete] $\C$-coloured traces.
Two states $s$ and $t$ are \emph{(divergence blind) branching
  bisimulation equivalent}, notation $s\Cc t$, if there exists a
consistent colouring $\C$ such that $\C(s)=\C(t)$.

They are \emph{divergence sensitive branching bisimulation equivalent},
notation $s\Ccl t$, if there exists a fully consistent colouring $\C$ such
that $\C(s)=\C(t)$.

A consistent colouring \emph{preserves divergence} if two states of the
same colour always have the same divergent $\C$-coloured traces.
Two states $s$ and $t$ are \emph{branching bisimulation equivalent
  with explicit divergence}, notation $s\Ccd t$, if there exists a
consistent, divergence preserving colouring $\C$ with $\C(s)=\C(t)$.
\end{defi}

\begin{figure*}[htb!]\centering
$
    \begin{array}{@{}ccccc@{\!\!}}
    a)\hspace{-0.4cm}
    \xymatrix@R=0.9cm@C=0.4cm{
            & *++[o][F][F*:mygray1]{s}  \ar[d]^{\mbox{$\tau$}}\ar@/_/[ld]_{\mbox{$\tau$}}\ar@/^/[rd]^{\mbox{$\tau$}} & \\
          *++[o][F][F*:mygray1]{t}\ar@(ld,lu)[]^{\mbox{$\tau\!$}} \ar[d]_{\mbox{$a$}} & *++[o][F][F*:mygray1]{u} \ar[d]_{\mbox{$a$}}& *++[o][F][F*:mygray1]{v}  \ar[d]_{\mbox{$a$}}\\
          *++[o][F][F*:mygray2]{x}  & *++[o][F][F*:mygray2]{y}  & *++[o][F][F*:mygray2]{z}  \ar@(rd,ru)[]_{\mbox{$\!\tau$}}
          }
    &  &
    b)\hspace{-0.4cm}
   \xymatrix@R=0.9cm@C=0.4cm{
            & *++[o][F][F*:mygray3]{s}  \ar[d]^{\mbox{$\tau$}}\ar@/_/[ld]_{\mbox{$\tau$}}\ar@/^/[rd]^{\mbox{$\tau$}} & \\
          *++[o][F][F*:mygray4]{t}\ar@(ld,lu)[]^{\mbox{$\tau\!$}} \ar[d]_{\mbox{$a$}} & *++[o][F][F*:mygray1]{u} \ar[d]_{\mbox{$a$}}& *++[o][F][F*:mygray1]{v}  \ar[d]_{\mbox{$a$}}\\
          *++[o][F][F*:mygray2]{x}  & *++[o][F][F*:mygray2]{y}  & *++[o][F][F*:mygray2]{z}  \ar@(rd,ru)[]_{\mbox{$\!\tau$}}
          }
  & &
    c)\hspace{-0.4cm}
  \xymatrix@R=0.9cm@C=0.4cm{
            & *++[o][F][F*:mygray3]{s}  \ar[d]^{\mbox{$\tau$}}\ar@/_/[ld]_{\mbox{$\tau$}}\ar@/^/[rd]^{\mbox{$\tau$}} & \\
          *++[o][F][F*:mygray4]{t}\ar@(ld,lu)[]^{\mbox{$\tau\!$}} \ar[d]_{\mbox{$a$}} & *++[o][F][F*:mygray1]{u} \ar[d]_{\mbox{$a$}}& *++[o][F][F*:mygray6]{v}  \ar[d]_{\mbox{$a$}}\\
          *++[o][F][F*:mygray2]{x}  & *++[o][F][F*:mygray2]{y}  & *++[o][F][F*:mygray5]{z}  \ar@(rd,ru)[]_{\mbox{$\!\tau$}}
          }
     \end{array}$

 \caption{Difference between a) $\bbisim$, b) $\dsbbisim$, and c) $\bbisimd$.}
 \label{fig::bb-example}
\end{figure*}

\noindent The difference between $\bbisim$, $\dsbbisim$, and $\bbisimd$ is
illustrated in the following example.

\begin{exa}\rm
Consider first the LTS and its colouring depicted in
Figure~\ref{fig::bb-example}a. This colouring is consistent and we
have $s\Cc t\Cc u\Cc v$ and $x\Cc y \Cc z$.
It is not fully consistent because state $t$ has a complete trace
  $\xymatrix@R=1cm@C=0.3cm{*++[o][F*:mygray1]{}}$
whereas $u$ has not. It is easy to see that every fully consistent
colouring must assign
different colours to states $t$ and $u$, and so that $t\,\not\!\Ccl u$.
One such colouring is given in Figure~\ref{fig::bb-example}b
and it shows that $u\Ccl v$ and $x\Ccl y \Ccl z$. Note, however, that this
colouring, although fully consistent, does not preserve divergence. State
$v$ has a divergent trace $\xymatrix@R=1cm@C=0.02cm{*++[o][F*:mygray1]{} & a
&*++[o][F*:mygray2]{} }$ whereas $u$ has not, and similarly state $z$ has a
divergent trace $\xymatrix@R=1cm@C=0.3cm{*++[o][F*:mygray2]{}}$ whereas $y$ has
not. Any colouring that preserves divergence must assign different colours to
states $u$ and $v$ and to states $y$ and $z$, meaning that $u\,\not\!\Ccd v$ and
$y\,\not\!\Ccd z$. One such colouring is given in Figure~\ref{fig::bb-example}c.
It shows that $x \Ccd y$. In fact, these are the only two (different)
states that are branching bisimulation equivalent with explicit divergence.
\end{exa}

\noindent
In the definition of $\Ccd$ above, consistency and preservation of divergence
appear as two separate properties of colourings. Instead we could have
integrated them by adding an extra bit ($\Delta$) at the end of those finite
coloured traces that stem from infinite paths. Likewise, $\Ccl$ could have been
defined by adding such an extra bit at the end of those finite coloured traces
that stem from maximal paths.

Lemmas~\ref{2} and~\ref{1} about colourings on Kripke structures
apply to labelled transition systems as well. The proofs are essentially
the same.

\begin{lem}\label{BB2}
Let $\C$ be a colouring such that two states with the same colour have
the same $\C$-coloured traces of length three (i.e.\ colour - action - colour).
Then $\C$ is consistent.\qed
\end{lem}

\begin{lem}\label{BB1}
Let $\C$ be a consistent colouring such that two states with the same
colour have the same complete $\C$-coloured traces of length one. Then $\C$
is fully consistent.\qed
\end{lem}

\begin{lem}\label{BBD}
Let $\C$ be a consistent colouring such that two states with the same
colour have the same divergent $\C$-coloured traces of length one. Then $\C$
preserves divergence.
\end{lem}
\begin{proof}
Exactly like the proof of Lemma~\ref{1}, but letting $\sigma$ be a
\emph{divergent} $\C$-coloured trace of $s$; $\pi,\pi'$ \emph{infinite}
paths; $\C(t')$ a \emph{divergent} $\C$-coloured trace of $t'$; and
$\rho',\rho''$ \emph{infinite} paths.
\end{proof}

\noindent
Branching bisimulation equivalence and branching bisimulation
equivalence with explicit divergence were originally defined in Van
Glabbeek \& Weijland \cite{GW96}.  There, only \emph{finite} coloured
traces are considered, and a consistent colouring was defined as a
colouring with the property that two states have the same colour only
if they have the same finite coloured traces.  By Lemma~\ref{BB2} this
yields the same concept of consistent colouring as
Definition~\ref{colourings} above.

In \cite{GW96}, a consistent colouring is said to \emph{preserve
divergence} if no divergent state has the same colour as a
nondivergent state. Here a state $s$ is \emph{divergent} if it is the
starting point of an infinite path of which all nodes have the same
colour. This is the case if $s$ has a divergent coloured trace of
length one. Now Lemma~\ref{BBD} says that the definition of
preservation of divergence from \cite{GW96} agrees with the one
proposed above. Hence the concepts of branching bisimulation and
branching bisimulation with explicit divergence of \cite{GW96} agree
with ours.

\begin{thm}\label{equivalence}
$\Cc$, $\Ccl$ and $\Ccd$ are equivalence relations.
\end{thm}
\begin{proof}
We show the proof for $\Cc$; the other two cases proceed likewise.

We will regard any equivalence relation on $S$ as a colouring, the
colour of a state being its equivalence class. Conversely, any
colouring can be considered as an equivalence relation on states.

The diagonal on $\States$ (i.e., the binary relation
$\{(\states,\states)\mid\states\in\States\}$) is a consistent
colouring, so $\Cc$ is reflexive.  That $\Cc$ is symmetric is
immediate from the required symmetry of colourings.

To prove that $\Cc$ is transitive, suppose $s\Cc t$ and $t\Cc u$.
So there exist consistent colourings $\C$ and $\D$ with
$\C(s)=\C(t)$ and $\D(t)=\D(u)$.  Let $\E$ be the finest equivalence
relation containing $\C$ and $\D$. Then $\E(s)=\E(t)=\E(u)$. It
suffices to show that $\E$ is consistent.

First of all note that the $\E$-colour of a state is completely
determined by its $\C$-colour, as well as by its $\D$-colour:
$\C(p)=\C(q) \Rightarrow \E(p)=\E(q)$ and
$\D(p)=\D(q) \Rightarrow \E(p)=\E(q)$ for all $p,q\in S$.
Thus, if two states have the same sets of $\C$-coloured traces or
the same sets of $\D$-coloured traces, they must also have the same
sets of $\E$-coloured traces.

Suppose $\E(p)=\E(q)$. Then there must be a sequence of states
$(p_i)_{0\leq i \leq n}$ such that $p=p_0$, $q=p_n$ and for $0\leq i <
n$ we have either $\C(p_i)=\C(p_{i+1})$ or $\D(p_i)=\D(p_{i+1})$.  As
$\C$ and $\D$ are consistent colourings, $p_i$ and $p_{i+1}$ have the
same $\C$-coloured traces or the same $\D$-coloured traces. In either
case they also have the same $\E$-coloured traces. This holds for
$0\leq i < n$, so $p$ and $q$ have the same $\E$-coloured traces.
Thus $\E$ is consistent.
\end{proof}

\begin{lem}\label{maximal}
Let $\C$ be a consistent colouring and $s\in S$. Then the complete
$\C$-coloured traces of $s$ consist of the $\C$-coloured traces of $s$
that are infinite, divergent, or maximal, in the sense that they
cannot be extended.
\end{lem}
\begin{proof}
By definition, infinite and divergent $\C$-coloured traces of $s$ are
complete. Let $\sigma$ be a maximal $\C$-coloured trace of $s$, and
let $\pi$ be a path from $s$ such that $\C(\pi)=\sigma$. Let $\pi'$ be
an extension of $\pi$ to a maximal path. As $\sigma$ is a maximal
$\C$-coloured trace, in the sense that it cannot be extended, we have
$\C(\pi')=\sigma$. Hence $\sigma$ is a complete $\C$-coloured trace of $s$.

Now let $\sigma$ be a complete $\C$-coloured trace of $s$ that is not
infinite, nor a divergent $\C$-coloured trace of $s$. In that case
$\sigma=\C(\pi)$ for $\pi$ a finite maximal path from $s$.  Let $t$ be
the last state of $\pi$. We have $\neg\exists a,t'.\ t\step{a}t'$.
Suppose, towards a contradiction, that $\sigma$ is not maximal, i.e.\
there is a path $\pi'$ from $s$ such that $\C(\pi')$ is a proper
extension of $\sigma$. Then there must be a state $u$ on $\pi'$ with
$\C(u)=\C(t)$, such that $u$ has a coloured trace $\sigma'$ of length
$>1$, which is a suffix of $\C(\pi')$.  As $\C$ is consistent,
$\sigma'$ is also a coloured trace of $t$, contradicting $\neg\exists
a,t'.\ t\step{a}t'$.
\end{proof}

\noindent
As for Kripke structures, for any colouring $\C$, the $\C$-coloured
traces of a state $s$ are the prefixes of the complete $\C$-coloured
traces of $s$.  Moreover, Lemma~\ref{maximal} says that the complete
$\C$-coloured traces of a state $s$ are completely determined by the
$\C$-coloured traces of $s$ together with the divergent
$\C$-coloured traces of $s$.  Hence, any colouring that is consistent
and preserves divergence is also fully consistent. Therefore, $\Ccd$
is finer than $\Ccl$, which is finer than $\Cc$.

The difference between $\Ccl$ and $\Ccd$ is that only the latter sees
the difference between those maximal finite coloured traces that stem
from finite paths (ending in \emph{deadlock}) and those that stem from
infinite paths (ending in \emph{livelock}).  For \emph{deadlock-free}
LTSs (having no states $s$ with $\neg\exists a,s'.\ s\step{a}s'$) the
equivalences $\Ccl$ and $\Ccd$ coincide.

\section{Translating between Kripke structures and labelled transition systems}
\label{translating}

\noindent
We presuppose a set $\Acts$ of \emph{actions} with a special element
$\sact\in\Acts$, and a set $\Props$ of \emph{atomic propositions}.
A \emph{doubly labelled transition system} (\dlts) is a structure
  $(\States,\Lab,\stepsym)$
that consists of
  a set of states $\States$,
  a \emph{labelling function}
    $\Lab:\States\rightarrow\Pow{\Props}$
and
  a \emph{labelled transition relation}
    $\stepsym\subseteq\States\times\Acts\times\States$.
From an {\dlts} one naturally obtains an LTS by omitting the labelling
function $\Lab$, and a Kripke structure by replacing the labelled
transition relation by one from which the labels are omitted. We call
these the LTS or Kripke structure \emph{associated} to the \dlts.
An {\dlts} $(\States,\Lab,\stepsym)$ is
\emph{consistent} if it satisfies the following three conditions:
\leftmargini 25pt
\begin{enumerate}\itemsep=0pt
\renewcommand{\theenumi}{\roman{enumi}}
\renewcommand{\labelenumi}{(\theenumi)}
\item \label{cnd:consistent1}
  if $\states\step{\acta}\statet$, then
  ($\Lab(\states)=\Lab(\statet)$ iff $\acta=\sact$);
\item \label{cnd:consistent2}
  if $\states\step{\acta}\statet$,
         $\state{s'}\step{\act{a}}\statet'$ and
         $\Lab(\states)=\Lab(\state{s'})$,
      then $\Lab(\statet)=\Lab(\state{t'})$; and
\item \label{cnd:consistent3}
  if $\states\step{\acta}\statet$,
         \plat{$\state{s'}\step{\act{b}}\statet'$},
         $\Lab(\states)=\Lab(\state{s'})$ and
         $\Lab(\statet)=\Lab(\state{t'})$,
      then $\acta=\act{b}$.
\end{enumerate}
Consistent {\dlts}s were introduced in De Nicola \& Vaandrager
\cite{DV95} for studying relationships between notions defined for
Kripke structures and notions defined for LTSs.
Condition~(\ref{cnd:consistent1}) states that a transition is
unobservable in the underlying Kripke structure (i.e., a transition
between states with the same label) if and only if it is an
unobservable transition in the underlying labelled transition system
(i.e., a $\silent$-transition).
Condition~(\ref{cnd:consistent2}) expresses that the label of the target
state of a transition is completely determined by the label of the
source state and the label of the transition. Consequently, the label
of a state $t$ reachable from a state $s$ is completely determined by
the label of $s$ and the sequence of labels of the transitions leading
from $s$ to $t$.
Condition~(\ref{cnd:consistent3}) says that the label of a transition is
fully determined by the labels of its source and target state.

\begin{exa}\rm
The three {\dlts}s from Figure~\ref{fig::lt2s}a are not consistent
because they violate conditions (i), (ii), and (iii), respectively; the
{\dlts} in Figure~\ref{fig::lt2s}b is consistent.
\end{exa}

\begin{figure*}[!ht]\centering
$
    \begin{array}{r@{\qquad}ccc@{\qquad}rc}
    a) & \xymatrix@R=1cm@C=0.3cm{
           *++[o][F]{} \newslabr{p} \ar@(u,l)[]_{\mbox{$a$\hspace{-1mm}}}
         }
       & \xymatrix@R=1cm@C=0.3cm{
          & *++[o][F]{} \newslabr{p} \ar[dl]_{\mbox{$a$}}
            \ar[dr]^{\mbox{$a$}} \\
          *++[o][F]{} \newslabl{q} && *++[o][F]{} \newslabr{r}
         }
       & \xymatrix@R=1cm@C=0.3cm{
          & *++[o][F]{} \newslabr{p} \ar[dl]_{\mbox{$a$}}
            \ar[dr]^{\mbox{$b$}} \\
          *++[o][F]{} \newslabl{q} && *++[o][F]{} \newslabr{q}
         }
       &
    b) & \xymatrix@R=1cm@C=0.8cm{
           & *++[o][F]{} \newslabr{p} \ar[dl]_{\mbox{$a$}}
             \ar[d]_{\mbox{$a$}} \ar[dr]^{\mbox{$b$}} \\
           *++[o][F]{} \newslabl{q}\ar[r]^{\mbox{$\tau$}} &
           *++[o][F]{} \newslabr{q} & *++[o][F]{} \newslabr{r}
         }
    \end{array}
$
 \caption{a) Three inconsistent {\dlts}s and b) a consistent \dlts.}
 \label{fig::lt2s}
\end{figure*}

\noindent Many semantic equivalences on LTSs, such as $\Cc$, $\Ccl$ and $\Ccd$,
are considered in the literature; for an overview see \cite{Gla93}.

\begin{defi}\rm\label{extending}
Any semantic equivalence $\sim$ on LTSs extends to {\dlts}s by
declaring, for all states $s$ and $t$ in an {\dlts}, that $s \sim t$
iff $\Lab(s)=\Lab(t)$ and $s \sim t$ in the associated LTS.

Any semantic equivalence $\sim$ on Kripke structures extends to {\dlts}s
by declaring, for all states $s$ and $t$ in an {\dlts}, that $s \sim
t$ iff $s \sim t$ in the associated Kripke structure.
\end{defi}

\noindent
The following theorem was proved in \cite{DV95} for finite consistent
{\dlts}s. Here we drop the finiteness restriction.

\begin{thm}\label{agreement}
On a consistent L$^2\!$TS, $\dbs$ equals $\Cc$, and $\s$ equals $\Ccl$.
\end{thm}
\begin{proof}
Suppose $s \dbs t$ [or $s \s t$]. Then there is a colouring $\C$ on
the states of the {\dlts} that is [fully] consistent on the associated
Kripke structure $\KS$ and satisfies $\C(s)=\C(t)$. By definition, this
entails $\Lab(s)=\Lab(t)$.  It remains to show that $\C$ is [fully]
consistent on the associated LTS L\@.  So let $\C(p)=\C(q)$, and let
$\sigma$ be a [complete] coloured trace of $p$ in L\@.  Using
symmetry, it suffices to show that $\sigma$ is also a [complete]
coloured trace of $q$ in L\@.  Let $\rho$ be obtained by omitting all
actions from the alternating sequence of states and actions $\sigma$.
Using direction ``only if'' of clause (i) in the definition of a
consistent {\dlts}, $\rho$ must be a [complete] coloured trace of $p$
in $\KS$.  As $\C$ is [fully] consistent on $\KS$, $\rho$ must also be a
[complete] coloured trace of $q$ in $\KS$.  Finally, using clauses (i)
``only if'' and (iii), $\sigma$ must be a [complete] coloured trace of
$q$ in L\@.

Now suppose $s \Cc t$ [or $s \Ccl t$]. Then $\Lab(s)=\Lab(t)$ and
there is a colouring $\C$ on the states of the {\dlts}, with
$\C(s)=\C(t)$, that is [fully] consistent on L\@. Let $\D$ be the
colouring given by $\D(p) := (\C(p),\Lab(p))$ for all $p\in S$, so
that $\D(p)=\D(q)\Leftrightarrow[\C(p)=\C(q) \wedge
\linebreak[2]
\Lab(p)=\Lab(q)]$. It suffices to show that $\D$ is [fully] consistent
on $\KS$.  The requirement $\D(p)=\D(q)\Rightarrow\Lab(p)=\Lab(q)$ is
built into the definition of $\D$. So let $\D(p)=\D(q)$, and let $\nu$
be a [complete] $\D$-coloured trace of $p$ in $\KS$.
Using symmetry, it suffices to show that $\nu$ is also a [complete]
$\D$-coloured trace of $q$ in $\KS$.
Using clause (i) ``only if'', there must be a [complete] $\D$-coloured
trace $\rho$ of $p$ in L such that $\nu$ is obtained from $\rho$
by omitting its actions.
Let $\sigma$ be the [complete] $\C$-coloured trace of $s$ in $\KS$ obtained
from $\rho$ by omitting the second component of each $\D$-colour of a
state.
As $\C(p)=\C(q)$ and $\C$ is [fully] consistent on L, $\sigma$ must
also be a [complete] $\C$-coloured trace of $q$ in L\@.
By applying clauses (i) ``if'' and (ii) one derives that $\rho$
is a [complete] $\D$-coloured trace of $q$ in L\@.
Therefore, again using clause (i) ``only if'', $\nu$ must be a
[complete] $\D$-coloured trace of $q$ in $\KS$.
\end{proof}

\begin{obs}\label{encoding}
For every Kripke structure $\KS$ there exists a consistent
L$^2\!$TS
{\rm D} such that $\KS$ is the Kripke structure associated to {\rm D}.
\end{obs}
\noindent
One way to obtain D is to label any transition $s \step{} t$ by the
pair $(\Lab(s),\Lab(t))$ (or simply by $\Lab(t)$) when
$\Lab(s)\neq\Lab(t)$, or $\tau$ when $\Lab(s)=\Lab(t)$. An alternative
is the label $(\Lab(s)-\Lab(t),\Lab(t)-\Lab(s))$, where
$(\emptyset,\emptyset)$ is identified with $\tau$.

Unlike the situation for Kripke structures (Observation~\ref{encoding})
it is not the case that every LTS can be obtained as the LTS
associated to a consistent \dlts. A simple counterexample is presented
in \cite{DV95}. Thus, in encoding LTSs as {\dlts}s, it is in general
not possible to keep the set of states the same.

\begin{defi}\rm\label{transformation}
An \emph{LTS-to-L$^2\!$TS transformation}
$\eta$ consist of a function
taking any LTS L to a consistent {\dlts} $\eta({\rm L})$, and in
addition taking any state $s$ in L to a state $\eta(s)$ in $\eta({\rm
L})$.  Such a transformation should at least satisfy $s \Ccl t
\Leftrightarrow \eta(s) \Ccl \eta(t)$, that is, it \emph{preserves}
(``$\Rightarrow$'') and \emph{reflects} (``$\Leftarrow$'') divergence
sensitive branching bisimulation equivalence, and likewise for $\Cc$,
and $\Ccd$.
\end{defi}
\noindent
A common LTS-to-{\dlts} transformation is presented in \cite{DV95}.
It takes an LTS ${\rm L}=(S,\rightarrow)$ to an {\dlts} $\eta({\rm L})$
by inserting a new state halfway along any transition $s\step{a}t$ with
$a \neq \tau$.  This new state is labelled $\{a\}$, and each of the
two transitions replacing $s\step{a}t$ (from $s$ to the new state and
from there to $t$) is labelled $a$. Transitions $s\step{\tau}t$ are
untouched. One takes $\eta(s)=s$ for $s\in S$ and all such states from
L are labelled with the same dummy symbol
in $\eta({\rm L})$. (Consult \cite{DV95} for
the formal definition and examples.) In \cite{DV95} it is shown that
this transformation preserves and reflects $\Ccl$; the same proof
applies to $\Cc$ and $\Ccd$.

An LTS-to-{\dlts} transformation $\eta$ yields an
LTS-to-Kripke-structure transformation that we also call $\eta$,
namely the one transforming an LTS L into the Kripke structure
associated to $\eta({\rm L})$. In fact, using Theorem~\ref{agreement}
and Observation~\ref{encoding}, any LTS-to-Kripke-structure
transformation $\eta$ that preserves and reflects the required
equivalences can be obtained in this way.

An LTS-to-{\dlts} transformation $\eta$ makes it possible to define
when a state $s$ in an LTS satisfies a {\CTLSminX} formula $\sfrmf$.
Namely, one defines $s \sat^\eta \sfrmf$ iff $\eta(s)\sat\sfrmf$.
This way, {\CTLSminX} can be used as temporal logic on LTSs.

\begin{thm} Let $s$ and $t$ be states in an LTS, and let $\eta$ be
an LTS-to-L$^2\!$TS transformation. Then\vspace{-1.7ex}
\begin{center}
$s \Cc t$ iff $s\models^\eta_{\it db} \sfrmf\Leftrightarrow
t\models^\eta_{\it db}\sfrmf$ for all \CTLSminX{} state formulas $\sfrmf$
\\
$s \Ccl t$ iff $s\models^\eta \sfrmf \Leftrightarrow t\models^\eta \sfrmf$
for all \CTLSminX{} state formulas $\sfrmf$.
\end{center}
\end{thm}
\begin{proof}
This is an immediate  consequence of the requirement that $\eta$
preserves and reflects $\Cc$ and $\Ccl$, in combination with
Theorems~\ref{dbs char},~\ref{stuttering char} and~\ref{agreement}.
\end{proof}

\section{Parallel composition} \label{sec:pcmp}

\noindent
For a behavioural equivalence to be useful in a process algebraic
setting, it is essential that it is a congruence for the operations
under consideration. In this section we prove that $\Ccd$ and $\Cc$
are congruences for the \emph{merge} or \emph{interleaving operator}
$\pcmpsym$.  This operator is often used to represent (asynchronous)
parallel composition.  However, $\Ccl$ fails to be a congruence for
$\pcmpsym$.  We characterise the least discriminating congruence that
makes all the distinctions of $\Ccl$ as $\Ccd$. In the following
definition we provide the necessary and sufficient conditions for a
binary operation \emph{on} the set of states of an LTS to qualify as a
merge.

\begin{defi}\rm
  A binary operation $\pcmpsym$ on the states of an LTS is a
  \emph{merge} if for all $\states,\statet,\stateu\in\States$ and for
  all $\acta\in\Acts$ it holds that
  $\states\pcmp\statet\step{\acta}\stateu$ \IFF{}
\begin{list}{$-$}{\leftmargin 18pt
                  \itemsep 0pt \parsep 0pt
                        \labelwidth\leftmargini\advance\labelwidth-\labelsep}
  \item there exists $\state{s'}\in\States$ such that
    $\states\step{\acta}\state{s'}$ and
    $\stateu=\state{s'}\pcmp\statet$; or
  \item there exists $\state{t'}\in\States$ such that
    $\statet\step{\acta}\state{t'}$ and
    $\stateu=\states\pcmp\state{t'}$.
\vspace{-5pt}
\end{list}
\end{defi}
\noindent
The structural operational semantics of any process calculus that
includes an operation for pure interleaving generates an LTS with
merge. Moreover, any LTS can be augmented to an LTS with merge, for
instance through a transition system specification \cite{AFV01} that
includes all states of the original LTS as constants and a binary
operation $\|$ with the usual structural operational rules for
interleaving parallel composition.
Henceforth we deal with LTSs with a merge $\|$.

\begin{thm}
  The relation $\Ccd$ is a congruence for $\|$,
  i.e., if $\states\bbisimd\!\statet$ and
  $\stateu\bbisimd\!\statev$, then
  $\states\pcmp\stateu\bbisimd\statet\pcmp\statev$.
\end{thm}
\begin{proof}
  Let $\brelsym$ be the reflexive and transitive closure of the
  relation
  \[\qquad
    \{(p\| q,\,p'\| q')\mid p\bbisimd p'\ \&\ q \bbisimd q'\}
  \enskip.
  \]
  Let $\C$ be the function that assigns to each state its equivalence
  class with respect to $\brel$. It suffices to prove that $\C$ is a
  consistent divergence preserving colouring.  So suppose
  $\C(r)=\C(r')$. Using Lemmas~\ref{BB2} and \ref{BBD} it suffices to
  show that $r$ and $r'$ have the same $\C$-coloured traces of length
  three and the same divergent $\C$-coloured traces of length one.
  It is straightforward, but notationally cumbersome, to establish
  this in the special case that $r=p\pcmp q$ and $r'=p'\pcmp q'$ with
  $p\bbisimd p'$ and $q\bbisimd q'$. The general case then follows
  by induction on the length of a chain of pairs from the relation
  displayed above showing that the pair $(r,r')$ is in its
  reflexive and transitive closure.
\end{proof}

\noindent
A similar proof shows that also $\Cc$ is a congruence for $\|$.
However, $\Ccl$ is not.

\begin{exa}\rm\label{deadlock}
Consider an LTS with merge that contains
  a state $0$ without outgoing transitions,
  a state $\Delta 0$ with a $\tau$-loop (an outgoing $\tau$-labelled
  transition to itself) and no other outgoing transitions, and
  a state $a$ with $a\step{a}0$ and no other outgoing transitions.
(Note that such an LTS is, e.g., generated by the structural
operational semantics of CCS with recursion.)
Then $0\Ccl \Delta 0$.
Figure~\ref{fig::congruence-example}a shows the fragment consisting
of the states $0$, $\Delta 0$ and $a$ of the LTS under consideration.
Figure~\ref{fig::congruence-example}b shows a fragment where the merge
is applied.
Observe that $0\pcmp a \,\not\!\Ccl \Delta 0 \pcmp a$.
The reason is that $\Delta 0 \pcmp a$ has a maximal path that stays in its
initial state, whereas $0\pcmp a$ has not. This problem does not apply to $\Cc$
because $0\pcmp a \Cc \Delta 0 \pcmp a$. It does not apply to $\Ccd$ because $0
\,\not\!\Ccd \Delta 0$.
\end{exa}

\begin{figure}\centering
$
\begin{array}{cccccccccccccccccc}
  a) \!\! \xymatrix@R=1cm@C=0.3cm{
          && *++[o][F]{\mbox{$a$}} \ar[d]^{\mbox{$a$}} \\
            *++[o][F]{\mbox{$\Delta 0$}}\ar@(r,u)[]_{\mbox{$\tau$}} & \Ccl &*++[o][F]{0}
    }
    & \qquad\quad &
    b)\quad
    \xymatrix@R=1cm@C=0.3cm{
          *+[o][F]{0\|a} \ar[d]^{\mbox{$a$}}  & \not\!\Ccl
          & *+[o][F]{\Delta0\|a} \ar[d]^{\mbox{$a$}} \ar@(u,l)[]_{\mbox{$\tau$}}\\
          *+[o][F]{0\|0} && *+[o][F]{\Delta0 \| 0} \ar@(l,d)[]_{\mbox{$\tau$}}
           }
\end{array}$
 \caption{$\dsbbisim$ is not a congruence for parallel composition}
 \label{fig::congruence-example}
\end{figure}

\noindent The example above involves a deadlock state, namely $0$.  This is
unavoidable, as on LTSs without deadlock $\Ccl$ coincides with $\Ccd$ (cf.\
Section~\ref{branching}) and hence is a congruence for $\|$.

The standard solution to the problem of an equivalence $\sim$ failing
to be a congruence for a desirable operator $Op$ is to replace it by the
coarsest congruence for $Op$ that is included in $\sim$ \cite{Mi89}. Applying
this technique to the current situation, the coarsest congruence for $\pcmpsym$
included in $\Ccl$ turns out to be $\Ccd$.

\begin{thm}\label{coarsest}
$\Ccd$ is the coarsest congruence for $\pcmpsym$ that is included in
$\Ccl$.\footnote{Strictly speaking, we merely show that $\Ccd$ is the
    coarsest congruence for $\pcmpsym$ that is included in \plat{$\Ccl$} and
    satisfies the Fresh Atom Principle (FAP). This principle,
    described in \cite{vG05}, is satisfied by a semantic equivalence $\sim$ on
    LTSs when $\sim$ on an LTS L can always be obtained as the restriction
    of $\sim$ on any given larger LTS of which L is a subLTS, and whose
    transition labels may be drawn from a larger set of actions than
    those of L\@. FAP allows us to use the state $a$ that figures in
    the proof of Theorem~\ref{coarsest}, regardless of whether such a
    state, or the fresh action $a$, occurs in the given LTS or not.
    FAP is satisfied by virtually all semantic equivalences documented
    in the literature, and can be used as a sanity check for meaningful
    equivalences \cite{vG05}.}
\end{thm}

\begin{proof}
We have already seen that $\Ccd$ is a congruence for $\pcmpsym$, and
that it is included in $\Ccl$. To show that it is the coarsest, we
need to show that if $\sim$ is any congruence for $\pcmpsym$ that is
included in $\Ccl$, then $\sim$ is included in $\Ccd$.  So let $\sim$
be such a congruence and assume $s \sim t$. We need to show that $s
\Ccd t$.  Let $a$ be an action that does not occur in any path from
$s$ or $t$.  Since $\sim$ is a congruence for $\pcmpsym$, we have
$s\pcmp a \sim t\pcmp a$, where $a$ is the state from
Example~\ref{deadlock}.  As $\sim$ is included in $\Ccl$ we obtain
$s\pcmp a \Ccl t\pcmp a$. Let $\C$ be a fully consistent colouring
with $\C(s\| a) = \C(t\| a)$. Define the colouring $\D$ by $\D(p)
=\C(p\| a)$ for $p$ a state reachable from $s$ or $t$, and $\D(p)=p$
otherwise. Then $\D(s)=\D(t)$.  It suffices to show that $\D$ is
consistent and preserves divergence, implying $s \Ccd t$.

So suppose $\D(p)=\D(q)$ with $p\neq q$. Then $\C(p\|a)=\C(q\|a)$.

First we show that $p$ and $q$ have the same $\D$-coloured traces.
Let $\sigma$ be a $\D$-coloured trace of $p$. Then $\sigma$ is also a
$\C$-coloured trace of $p\pcmp a$.  As $p\pcmp a$ and $q\pcmp a$ have
the same complete $\C$-coloured traces, they surely have the same
$\C$-coloured traces (for the coloured traces of a state are the
prefixes of its complete coloured traces). Hence $\sigma$ is a
$\C$-coloured trace of $q\pcmp a$.  As $p$ is reachable from $s$ or
$t$, the action $a$ cannot occur in $\sigma$. Therefore, $\sigma$ must
also be a $\D$-coloured trace of $q$. By symmetry, any $\D$-coloured
trace of $q$ is also a $\D$-coloured trace of $p$, and hence $p$ and
$q$ have the same $\D$-coloured traces.

Next, we show that $p$ and $q$ have the same divergent
$\D$-coloured traces. So let $\sigma$ be a divergent $\D$-coloured
trace of $p$. Then $\sigma$ is also a divergent $\C$-coloured trace of
$p\pcmp a$. Hence $\sigma$ is a complete $\C$-coloured trace of $p \pcmp
a$ and thus also of $q\pcmp a$. As the action $a$ cannot occur in
$\sigma$, it is not possible that $\sigma$ stems from a finite maximal
path from $q\pcmp a$. Therefore, $\sigma$ must be a divergent
$\C$-coloured trace of $q\pcmp a$, and hence a divergent $\D$-coloured
trace of $q$. Again invoking symmetry, $p$ and $q$ have the same
divergent $\D$-coloured traces.

It follows that $\D$ is consistent and preserves divergence; thus $s \Ccd t$.
\end{proof}

\noindent
So if one is in search of a semantics such that, for $s$ and
$t$ states in an LTS,
\begin{list}{$-$}{\leftmargin 18pt \topsep 0pt \parsep 0pt \itemsep 0pt
                  \labelwidth\leftmargini\advance\labelwidth-\labelsep}
\item if there is a {\CTLSminX} state formula $\sfrmf$ such that
  $s \sat^\eta \sfrmf$ but $t \not\sat^\eta \sfrmf$, then $s$ and $t$
  should be distinguished,
\item if $s$ and $t$ can be distinguished after placing them in a
  context $\_\!\_ \pcmp u$ for some $u$, then they should be
  distinguished to start with, and
\item no two states should be distinguished unless this is
  required by the previous two conditions,
\end{list}
then branching bisimulation semantics with explicit divergence is the
answer, for $s \Ccd t$ iff for all $u$ and all $\sfrmf \in \SFrm$ we
have $s\pcmp u \sat^\eta \sfrmf \Leftrightarrow t\pcmp u \sat^\eta \sfrmf$.

\section{Adding deadlock detection to \texorpdfstring{\CTLSminX{}}{CTL*X}}
\label{sec:newctls}

\noindent
We saw above that there are important properties of states $s$ in an
LTS that can be expressed in terms of a context $\_\!\_\pcmp u$ and a
{\CTLSminX} formula $\sfrmf$, namely as ${s \| u} \sat^\eta \sfrmf$, but
that cannot be directly expressed in terms of {\CTLSminX}.  This is
somewhat unsatisfactory, and therefore we propose an extension of
{\CTLSminX} in which this type of property can be expressed directly.
We add a path modality $\infty$ that is valid on a path $\pi$ iff
$\pi$ is infinite. This path modality, or actually an equally
expressive one, was studied prior by Kaivola \& Valmari \cite{KV92} in
the context of Linear Temporal Logic without the next state
operator---see Section~\ref{sec:LTL}.

\begin{defi}\rm
The syntax of {\CTLSD} is given by
\[
    \sfrmf ::=\
      \propp\ \mid\
      \compl\sfrmf\ \mid\
      \Meet\SFrm[\Phi']\ \mid\
      \is\pfrmp
\qquad\qquad
    \pfrmp ::=\
      \sfrmf\ \mid\
      \compl\pfrmp\ \mid\
      \Meet\PFrm[\Psi'] \mid\
      \pfrmp\Until\pfrmp \mid\
      \infty
\]
  with $\propp\in\Props$, $\sfrmf\in\SFrm$,
  $\SFrm[\Phi']\subseteq\SFrm$,
  $\pfrmp\in\PFrm$ and $\PFrm[\Psi']\subseteq\PFrm$.
\\
Validity is defined as in Definition~\ref{validity}, but adding the clause
\begin{list}{$-$}{\leftmargin 18pt
                        \labelwidth\leftmargini\advance\labelwidth-\labelsep}
  \item $\pthp\sat\infty$ \IFF{} the path $\pthp$ is infinite.
  \end{list}
\end{defi}
\noindent
We write $\exists^\infty\pfrmp$ for $\exists(\infty\wedge\pfrmp)$;
this formula holds in a state $s$ if there exists an infinite path
$\pthp$ from $s$ such that $\pthp\sat\pfrmp$. Likewise
$\forall^\infty\pfrmp = \forall(\infty\rightarrow\pfrmp)$ holds in $s$
if for all infinite paths $\pthp$ from $s$ we have that $s\sat\pfrmp$.
These constructs are \emph{dual}, in the sense that
$s\sat\neg\exists^\infty\pfrmp$ iff $s\sat\forall^\infty\neg\pfrmp$.

The negation of $\infty$ holds for a maximal path $\pthp$ iff $\pthp$
is finite, and hence ends in a deadlock. It is tempting to simply
extend $\CTLSminX$ with a state formula $\delta$ such that
$s\sat\delta$ iff $\neg\exists s'.\ s\step{} s'$. This would make it
possible to express $\infty$ as $\neg{\sf F}\delta$. However, this
would make the resulting logic too expressive: the two states in the
Kripke structure $\circ\step{}\circ$ (with the empty labelling) are
branching bisimulation equivalent with explicit divergence, yet they
would be distinguished by this extension of {\CTLSminX}, as only the
last state satisfies $\delta$.

{\CTLSD} is an extension of {\CTLSminX}. There is no need for a
similar extension of {\CTLS}, for $\delta$ can be expressed as
$\neg\exists{\sf X}\top$. In particular, {\CTLSD} is not more
expressive than {\CTLS}.

The definition of branching bisimulation equivalence with explicit
divergence lifts easily to Kripke structures: $s\Ccd t$, for
$s$ and $t$ states in a Kripke structure, iff there exists a
consistent and divergence preserving colouring $\C$ such that
$\C(s)=\C(t)$. Here \emph{divergence preserving} is defined as in
Section~\ref{branching};
by Lemma~\ref{BBD}, this time applied to Kripke structures,
a consistent colouring preserves divergence iff, for any states $s$ and $t$,
$\C(s)=\C(t)$ implies
\begin{center}
for any infinite path $\pi$ from $s$ with $\C(\pi)=\C(s)$\\
there is an infinite path $\rho$ from $t$ with $\C(\rho)=\C(t)$.
\end{center}

\begin{thm}\label{bbed char}
$s \Ccd t$ iff $s\models \sfrmf \Leftrightarrow t\models\sfrmf$
for all \CTLSD{} state formulas $\sfrmf$.
\end{thm}
\begin{proof}
``Only if'' goes as in the proof of Theorem~\ref{dbs char},
  reading $\sat$ for $\sat_{\it db}$, requiring $\C$ to be
  \emph{consistent and divergence preserving}, and, in the second
  paragraph, requiring the paths $\pi$ and $\rho$ to be maximal and
  $\C(\pi)$ to be a \emph{complete} coloured trace of $s$ and $t$.
  Here we use that if a colouring is consistent and divergence
  preserving, then two states with the same colour must also have the
  same complete coloured traces. This follows from Lemma~\ref{maximal},
  this time applied to Kripke structures.

  There is one extra case to check.
  Suppose $\C(\pi)=\C(\rho)$ and $\pi\sat\infty$, but $\rho\not\sat\infty$.
  Then the last state $t$ of $\rho$ has the same colour $\C(t)$ as one
  of the states $s$ of $\pi$. Let $\pi'$ be the (infinite) suffix of
  $\pi$ starting at $s$. Then $\C(\pi')=\C(s)=\C(t)$, yet there is no
  infinite path from $t$, contradicting that $\C$ is divergence preserving.

``If'' goes as in the proof of Theorem~\ref{dbs char}, but
  this time we also have to show that $\C$ preserves divergence.
  So let $s$ and $t$ be states and $\pi$ an infinite path from $s$ with
  $\C(\pi)=\C(s)=\C(t)=C$. Let
\begin{center}
$\mathcal{U}=\{u\mid$ there is a path from $t$ to $u$ and $\C(u)\neq C\}$.
\end{center}
For every $u\!\in \mathcal{U}$ pick a \CTLSD{} formula $\sfrmf_u
\in C-\C(u)$.  Now $s \models \exists^\infty {\sf G}(\bigwedge_{u\in
\mathcal{U}} \sfrmf_u)$ and, as $\C(s)=\C(t)$, also $t \models
\exists^\infty {\sf G} (\bigwedge_{u\in\mathcal{U}} \sfrmf_u)$.  Thus,
there is an infinite path $\rho$ from $t$ such that $t' \models
\bigwedge_{u\in \mathcal{U}} \sfrmf_u$ for all states $t'$ in
$\rho$. It follows that $t' \not\in \mathcal{U}$. Hence $\C(t')=C$ and
thus $\C(\rho)=C$.
\end{proof}

\section{Adding deadlock detection to \texorpdfstring{\CTLminX}{CTLX}}
\label{sec:newctl}

\noindent
{\CTLminX} is the sublogic of {\CTLSminX} that only allows path formulas
of the form $\sfrmf\Until\sfrmf'$ and $\neg(\sfrmf\Until\sfrmf')$,
where $\sfrmf$ and $\sfrmf'$ are state formulas.
Equivalently, it can be defined as only allowing path
formulas of the form $\sfrmf\Until\sfrmf'$ and ${\sf G}\sfrmf$,
for we have
$$s\sat\exists{\sf G}\sfrmf\mbox{~~iff~~}s\sat\exists\neg(\top\Until\neg\sfrmf)$$
$$s\sat\exists\neg(\sfrmf\Until\sfrmf')\mbox{~~iff~~}
  s\sat\exists[(\neg\sfrmf')\Until\neg(\sfrmf\vee\sfrmf')]
  \vee \exists{\sf G}\neg\sfrmf' \;.$$
Theorems~\ref{dbs char} and~\ref{stuttering char} are also valid when
using $\CTLminX$ instead of $\CTLSminX$, for their proofs use no other
temporal constructs than $\is(\sfrmf\Until\sfrmf')$ and $\is{\sf G}\sfrmf$.

A natural proposal for {\CTLD} would be to add the path quantifier
  $\exists^\infty$ to {\CTLminX}, thus yielding the syntax
\[
\begin{array}{rcl}
\sfrmf & ::= &
      \propp\ \mid\
      \compl\sfrmf\ \mid\
      \Meet\SFrm[\Phi']\ \mid\
      \exists(\sfrmf\Until\sfrmf) \mid\
      \exists^\infty(\sfrmf\Until\sfrmf) \mid\
      \exists{\sf G}\sfrmf \mid\
      \exists^\infty{\sf G}\sfrmf\ .
\end{array}
\]
However, we can economise on that, for
$$s\sat\exists^\infty(\sfrmf\Until\sfrmf') \mbox{~~iff~~}
  s\sat\exists(\sfrmf\Until(\sfrmf'\wedge\exists^\infty{\sf G}\top))$$
$$s\sat\exists{\sf G}\sfrmf \mbox{~~iff~~}
  s\sat\exists^\infty{\sf G}\sfrmf \vee
  \exists(\sfrmf\Until (\all{\sf G}\sfrmf))$$
where $\all{\sf G}\sfrmf$ is an abbreviation for $\neg\is(\top\Until\neg\sfrmf)$.
Hence {\CTLD} can be given by the syntax
$$  \sfrmf ::=\
      \propp\ \mid\
      \compl\sfrmf\ \mid\
      \Meet\SFrm[\Phi']\ \mid\
      \exists(\sfrmf\Until\sfrmf) \mid\
      \exists^\infty{\sf G}\sfrmf\ .
$$
It follows immediately from the proof of Theorem~\ref{bbed char}
that this language is sufficiently expressive to characterise
branching bisimulation equivalence with explicit divergence:

\begin{thm}
$s \Ccd t$ iff $s\models \sfrmf \Leftrightarrow t\models\sfrmf$
for all \CTLD{} formulas $\sfrmf$.\qed
\end{thm}

\noindent
It is tempting to simply write $\exists^\infty{\sf G}$ as
$\exists{\sf G}$; that is, to keep the same syntax as for {\CTLminX}
but define its semantics in such a way that $\is(\sfrmf\Until\sfrmf')$
asks merely for a finite path, whereas $\is{\sf G}\sfrmf$ asks for an
infinite one. This \emph{deadlock sensitive} interpretation of
{\CTLminX} is an alternative for the interpretation of \cite{DV95}.
It is consistent with the classical interpretation of {\CTL}
\cite{EH86,BCG88}, as for total Kripke structures there is no
difference between $\exists^\infty$ and $\exists$.

\section{The deadlock extension of Kripke structures}

\noindent
Following De Nicola \& Vaandrager \cite{DV95} we have applied
{\CTLSminX} to non-total Kripke structures by using maximal instead of
infinite paths in the definition of validity.  As remarked in
Section~\ref{CTL}, the same effect can be obtained by transforming a
non-total Kripke structure into a total one by adding a self-loop
$s\step{}s$ to every deadlock state $s$, and applying the standard
{\CTLSminX} semantics to the resulting total Kripke structure.
However, the latter does not apply to {\CTLSD}, because the self-loop
$s\step{}s$ invalidates the formula $\exists\neg\infty$ that holds in
any deadlock state $s$. Here we define another transformation on
Kripke structures that makes every Kripke structure total, and allows
the encoding of {\CTLSD} in terms of {\CTLSminX}.

\begin{defi}\rm \label{def:deadlockextension}
The \emph{deadlock extension} $\de(\KS)$ of a Kripke structure $\KS$ is
obtained by the addition of a fresh state $s_\delta$, labelled
by the fresh atomic proposition $\delta$, together with a transition
from $s_\delta$ and from every deadlock state in $\KS$ to $s_\delta$.
\end{defi}

\noindent An example of this transformation is depicted in
Figure~\ref{fig::deadlock-ext}.

\begin{figure}
 \centering
$ \begin{array}{cccccccc}
 \KS:&& \xymatrix@R=1cm@C=1cm{
          *++[o][F]{} \newslabr{p} \ar@(l,u)[]^{} \ar[d]_{}  \ar[r]_{} & *++[o][F]{} \newslabr{q} \\
          *++[o][F]{} \newslabr{r} &
          }
 &\qquad& \de(\KS):&& \xymatrix@R=1cm@C=1cm{
          *++[o][F]{} \newslabr{p} \ar@(l,u)[]^{} \ar[d]_{} \ar[r]_{} & *++[o][F]{} \newslabr{q} \ar[d]_{} \\
          *++[o][F]{} \newslabr{r} \ar[r]_{} & *++[o][F]{s_\delta} \slabr{\;\delta}\ar@(r,d)[]^{}
          }
 \end{array}
$
 \caption{Deadlock extension of a Kripke structure}
 \label{fig::deadlock-ext}
\end{figure}

\begin{thm}\label{deadlock extension}
Let $\KS$ be a Kripke structure, with states $s$ and $t$.
Then $s\Ccd t$ within the Kripke structure $\KS$ iff
$s\Ccd t$ within the Kripke structure $\de(\KS)$.
\end{thm}

\begin{proof}
``If'': Let $\D$ be a consistent and divergence preserving colouring
on $\de(\KS)$. Note that $\D(s_\delta)\neq\D(s)$ for any state $s\neq
s_\delta$ in $\de(\KS)$. Let $\C$ be the restriction of $\D$ to the
states of $\KS$. Then the $\C$-coloured traces of a state $s$ in $\KS$
equal the $\D$-coloured traces of $s$ in $\de(\KS)$, but with the
colour $\D(s_\delta)$ omitted from the end of such traces.  It follows
that $\C$ is consistent. It preserves divergence by Lemma~\ref{BBD}.

``Only if'': Let $\C$ be a consistent and divergence preserving
colouring on $\KS$. Extend it to a colouring $\D$ on $\de(\KS)$ by
assigning a fresh colour $\delta$ to the extra state $s_\delta$ of
$\de(\KS)$. It suffices to check that $\D$ is consistent and divergence
preserving.

{\it Claim.} From any state $s$ in $\KS$ with the same colour as a
deadlock state $t$ in $\KS$ there must be a path $\pi$ to a deadlock
state such that $\C(\pi)=\C(t)$.

{\it Proof of claim}. As $t$ has no $\C$-coloured traces of length
two, neither does $s$, and as $t$ has no divergent $\C$-coloured
traces, neither does $s$. Thus, all paths from $s$ are finite and only
pass through states with colour $\C(t)$.

{\it Application of the claim}. The $\D$-coloured traces of length two
of a state $s \neq s_\delta$ in $\de(\KS)$ are the $\C$-coloured traces
of length two of the state $s$ in $\KS$, together with the trace
$\C(t)\delta$ in case $s$ has the same colour as a deadlock state $t$
in $\KS$. Thus $\D$ is consistent by Lemma~\ref{2}, and preserves
divergence by Lemma~\ref{BBD}.
\end{proof}
The ``if''-direction of the theorem, with a similar proof, also
applies to $\s$ and $\dbs$, but the ``only if''-direction does not. As a
counterexample, let $\KS$ be a Kripke structure with a deadlock state
$d$ (having no outgoing transitions) and a livelock state $l$ (with a
self-loop as its only one outgoing transition); neither state satisfies
any atomic propositions. In $\KS$ we have $d \s l$, and hence $d \dbs l$,
but in $\de(\KS)$ we have $d \not\dbs l$, and hence $d \not\s l$.

Considering that Kripke structures of the form $\de(\KS)$ are total,
and that on total Kripke structures $\s$ and $\Ccd$ coincide, it is in
fact impossible to define a transformation like $\de$ for which
Theorem~\ref{deadlock extension} holds for both $\Ccd$ and $\s$.

Now let $\eta$ be an arbitrary LTS-to-\dlts-transformation, yielding
an LTS-to-Kripke-structure transformation that is also called $\eta$
(see Section~\ref{translating}). Then $\de \circ \eta$ is not a valid
LTS-to-Kripke-structure transformation as intended in \cite{DV95},
for it fails to preserve \mbox{$\Ccl/\s$} and \mbox{$\Cc/\dbs$} (cf.\
Definition~\ref{transformation}). Yet, it satisfies $$s \Ccd t
~~\Leftrightarrow~~ \de\circ\eta(s) \s \de\circ\eta(t)$$
(because $s \Ccd t ~\Leftrightarrow~ \eta(s) \Ccd \eta(t)
          ~\Leftrightarrow~ \de\circ\eta(s) \Ccd \de\circ\eta(t)$
and on total Kripke structures $\Ccd$ and $\s$ coincide), and as such
it is a suitable transformation for defining validity of {\CTLSminX}
formula on states in LTSs. We obtain:

\begin{cor}\label{modal char}
Let $s$ and $t$ be states in an LTS, and let $\eta$ be
an LTS-to-L$^2\!$TS transformation. Then
\vspace{-1.7ex}
\begin{center}
\mbox{}\hfill
$s \Ccd t$ iff $s\models^{\de\circ\eta} \sfrmf\Leftrightarrow
t\models^{\de\circ\eta}\sfrmf$ for all \CTLSminX{} state formulas $\sfrmf$.
\qed
\end{center}
\end{cor}

\noindent
Thus, one way to make {\CTLSminX} suitable for dealing with deadlock
behaviour on LTSs is to stick to total Kripke structures and
translate LTSs to Kripke structures by a translation $\de\circ\eta$
instead of a transformation $\eta$ as proposed in \cite{DV95}.
This way branching bisimulation equivalence with explicit divergence
becomes the natural counterpart of stuttering equivalence on Kripke
structures, and we have the modal characterisation of
Corollary~\ref{modal char}.

An alternative is to stick to more natural transformations $\eta$
meeting the criteria on Definition~\ref{transformation}, apply the
definition of validity of {\CTLSminX} formulas to non-total Kripke
structures as in \cite{DV95}, and extend {\CTLSminX} to {\CTLSD} as
indicated in Section~\ref{sec:newctls}.

Below we show that these solutions lead to equally expressive logics
on LTSs.

\begin{defi}\rm\label{deadlock encoding}
Given a set of atomic propositions, let {\CTLSd} be the logic
{\CTLSminX} extended with an extra atomic proposition
$\delta$. The mappings $\dec$ from {\CTLSD} to {\CTLSd} formulas and
$\enc$ from {\CTLSd} to {\CTLSD} formulas are defined inductively by
$$\begin{array}{@{}r@{~=~}l@{\qquad}r@{~=~}l@{}}
\dec (p) & p                       &
\enc\ (p) & p                       \\
\dec (\neg \sfrmf) & \neg\delta \wedge \neg \dec(\sfrmf) &
\enc (\neg \sfrmf) & \neg \enc(\sfrmf) \\
\dec (\bigwedge_{i\in I}\sfrmf_i) & \bigwedge_{i\in I} \dec(\sfrmf_i) &
\enc (\bigwedge_{i\in I}\sfrmf_i) & \bigwedge_{i\in I} \enc(\sfrmf_i) \\
\dec (\exists \pfrmp) & \exists \dec(\pfrmp) &
\enc (\exists \pfrmp) & \exists \enc(\pfrmp) \\
\dec (\neg \pfrmp) & \neg\delta \wedge \neg \dec(\pfrmp) &
\enc (\neg \pfrmp) & \neg \enc(\pfrmp) \\
\dec (\bigwedge_{i\in I}\pfrmp_i) & \bigwedge_{i\in I} \dec(\pfrmp_i) &
\enc (\bigwedge_{i\in I}\pfrmp_i) & \bigwedge_{i\in I} \enc(\pfrmp_i) \\
\dec (\pfrmp \Until  \pfrmp') & \dec(\pfrmp) \Until \dec(\pfrmp')&
\enc (\pfrmp \Until  \pfrmp') & (\enc(\pfrmp) \Until \delta_{\pfrmp'}) \vee
                                (\enc(\pfrmp) \Until \enc(\pfrmp'))\\
\dec (\infty) & \neg {\sf F}\delta &
\enc (\delta) & \neg \top.
\end{array}$$
Here $\delta_{\pfrmp'}=\left\{\begin{array}{l}\delta~\mbox{if}~s_\delta
\sat \exists \pfrmp'\\\neg\top~\mbox{otherwise}\end{array}\right.$,
and $\pfrmp \Until \delta$ abbreviates $\neg\infty \wedge {\sf G}
\pfrmp$.
\end{defi}
We remark that checking whether $s_\delta \sat \exists \pfrmp'$ is
simple: just substitute $\top$ for $\delta$ and $\bot$ for all other
atomic propositions in $\pfrmp'$, while simplifying subformulas
$\psi_1 \Until \psi_2$ to $\psi_2$. The latter is justified because
the unique infinite path starting from $s_\delta$ has only itself as suffix.

\begin{thm}\label{thm:equally expressive}
Let {\KS} be a Kripke structure and $s$ a state in {\KS}.
Then for any {\CTLSD} state formula $\sfrmf$ we have
 $s\sat\sfrmf$ in $\KS$ iff $s\sat\dec(\sfrmf)$ in $\de(\KS)$,
and for any {\CTLSd} state formula $\sfrmf$ we have
 $s\sat\sfrmf$ in $\de(\KS)$ iff $s\sat\enc(\sfrmf)$ in $\KS$.
\end{thm}
\begin{proof}
For a state formula $\sfrmf$, let $\Id{\sfrmf}_{\KS}$ denote the set
of states $s$ in $\KS$ with $s\sat\sfrmf$.  Likewise, for a path
formula $\pfrmp$, $\Id{\pfrmp}_{\KS}$ denotes the set of maximal paths
$\pi$ in $\KS$ with $\pi\sat\sfrmf$.
Note that there is a bijective correspondence between the maximal
paths in $\KS$ and those in $\de(\KS)$ not starting in
$s_\delta$. A straightforward structural induction shows that
$\Id{\sfrmf}_{\KS} = \Id{\dec(\sfrmf)}_{\de(\KS)}$ for any {\CTLSD}
state formula $\sfrmf$ and, up to the aforementioned bijective
correspondence, $\Id{\pfrmp}_{\KS} = \Id{\dec(\pfrmp)}_{\de(\KS)}$ for
any {\CTLSD} path formula $\pfrmp$.

For the second statement, let $\pi_\delta$ be the unique path in
$\de(\KS)$ starting in $s_\delta$.  A straightforward structural
induction shows that $\Id{\sfrmf}_{\de(\KS)} - \{s_\delta\} =
\Id{\enc(\sfrmf)}_{\KS}$ for any {\CTLSd} state formula $\sfrmf$ and,
up to the above bijective correspondence, $\Id{\pfrmp}_{\de(\KS)} -
\{\pi_\delta\} = \Id{\enc(\pfrmp)}_{\KS}$ for any {\CTLSd} path
formula $\pfrmp$.
\end{proof}

\noindent
In {\CTLSD} the path modality $\infty$ is equally expressive as the
path modality $\pfrmp\Until\delta$ of Definition~\ref{deadlock encoding},
saying of a path that it is finite and all its suffixes satisfy $\pfrmp$.
This is because
 $~\pi\sat\pfrmp\Until\delta
      ~\Leftrightarrow~
   \pi\sat\neg\infty\wedge{\sf G}\pfrmp$
and
  $\pi \sat \infty
      ~\Leftrightarrow~
   \pi \sat \neg {\sf F} \delta
      ~\Leftrightarrow~
   \pi \sat\neg \top \Until \delta$.
In this light, the encoding $\dec$ of $\CTLSD$ into $\CTLSd$ merely
adds a conjunct $\neg\delta$ here and there.  These conjuncts are not
optional; they enable, for instance, the correct translation of the
{\CTLSD} path formula ${\sf G}p$ by the {\CTLSd} formula
$\neg\delta\wedge {\sf G}(\delta \vee p)$.

Recall that in Section~\ref{sec:newctls} we considered extending
$\CTLSminX$ with a state formula $\delta$ such that $s\sat\delta$ iff
$\neg\exists s'.\ s\step{} s'$. We then argued that this would make
the resulting logic too expressive. Note that in our current proposal
the atomic proposition $\delta$ only holds in the fresh state
$s_\delta$ of the deadlock extension $\de(\KS)$ of a Kripke structure
$\KS$ and not in any of the original states of $\KS$. As a
consequence, in \CTLSD{}, which does not have the next state modality
{\sf X}, we can express the property
that deadlock is unavoidable (when all paths from an original state
of $\KS$ lead to deadlock), but we still cannot express the property
of \emph{being deadlocked} (i.e., the property that holds in an
original state of $\KS$ iff no further transitions are possible).

\begin{thm}
Also the logics $\CTLd$ and $\CTLD$ are equally expressive.
\end{thm}

\noindent\textit{Proof.\/}
This follows because $\dec$ can be restricted to a mapping from {\CTLD} to
{\CTLd} formula and $\enc$ to a mapping from {\CTLD} to {\CTLd} formula.
In particular,\vspace{1ex}
\begin{gather*}
  \dec (\exists(\sfrmf \Until  \sfrmf'))
 = \exists(\dec(\sfrmf) \Until \dec(\sfrmf'))
\qquad
  \dec (\exists {\sf G}^\infty\sfrmf)
 = \exists {\sf G} (\neg\delta\wedge\dec(\sfrmf))
\\[1ex]
  \enc (\exists(\sfrmf \Until \sfrmf'))
  = \left\{\begin{array}[c]{l@{\qquad}l@{}}
      \exists (\enc(\sfrmf) \Until \enc(\sfrmf')) \vee \exists
        (\enc(\sfrmf) \Until (\neg\exists^\infty{\sf G}\top \wedge
        \exists {\sf G}\enc(\sfrmf) ))
          & \mbox{if } s_\delta \sat \sfrmf'\\
      \exists (\enc(\sfrmf) \Until \enc(\sfrmf')) & \mbox{otherwise}\\
    \end{array}\right.
\intertext{and}
  \enc (\exists {\sf G}\sfrmf)
  = \left\{\begin{array}[c]{l@{\qquad}l@{\hspace{7.4cm}}r@{\hspace{-3pt}}}
      \exists {\sf G}^\infty \enc(\sfrmf)
        & \mbox{if } s_\delta \sat \sfrmf'
    \\
      \exists {\sf G} \, \enc(\sfrmf)
        & \mbox{otherwise}. & \qEd
    \end{array}\right.
\end{gather*}

\section{Linear temporal logic with deadlock detection}\label{sec:LTL}

\noindent
\emph{Linear Temporal Logic} \cite{Pn77} (\LTL{}) is the sublogic of
{\CTLS} that allows propositional variables $p \mathbin\in \Props$ but
no other state formulas to be used as path formulas.  Path formulas
are applied to states by an implicit universal quantification:
\mbox{$s \sat \psi$} iff $s \sat \forall\psi$.  In this section we
explore the programme of this paper in the setting of \LTLminX{}
(\LTL{} without the next state modality), and compare the results with
the branching time case.

First we characterise the equivalence induced on the states of a
Kripke structure $(S,\Lab,\stepsym)$ by validity of {\LTLminX}
formulas.
We can conveniently use the notion of complete coloured
traces in this characterisation, observing that $\Lab$ is a colouring
in the sense of Definition~\ref{stuttering}.
We write $s \ls t$ if the states $s$ and $t$ have the same complete
$\Lab$-coloured traces.
Now two states satisfy the same {\LTLminX} formulas iff they
have the same complete $\Lab$-coloured traces.

\begin{thm}\label{LTLchar}
$s \ls t$ iff $s\models \pfrmp \Leftrightarrow t\models\pfrmp$
for all \LTLminX{} formulas $\pfrmp$.
\end{thm}

\begin{proof}
``Only if'':
   Note that, to show that
     $s \ls t$ implies
        $s\models \pfrmp \Leftrightarrow t\models\pfrmp$,
   it suffices to prove that
     if $\Lab(\pthp)=\Lab(\pthr)$
     then $\pthp\models \pfrmp \Leftrightarrow \pthr\models\pfrmp$.
   We proceed by structural induction on $\pfrmp$.

   From $\Lab(\pthp)=\Lab(\pthr)$ it follows that the first states of
   $\pthp$ and $\pthr$ have the same colour, and hence if $\pfrmp=p$
   with $p\in\Props$ then $\pthp\models \pfrmp \Leftrightarrow \pthr\models\pfrmp$.
   The cases $\pfrmp=\neg\pfrmp'$ and $\pfrmp=\Meet\PFrm[\Psi']$ follow
   immediately from the induction hypothesis.

   Finally, let $\pfrmp=\pfrm{\psi'}\Until\pfrm{\psi''}$
   and suppose that
     $\pthp\models\pfrmp$.
   Then there exists a suffix $\pth{\pi'}$ of $\pthp$ such that
     $\pth{\pi'}\sat\pfrm{\psi''}$
   and
     $\pth{\pi''}\sat\pfrm{\psi'}$
       for all $\pthp\suff\pth{\pi''}\psuff\pth{\pi'}$.
   As $\Lab(\pthp)=\Lab(\pthr)$, there must be a suffix $\pth{\rho'}$ of
   $\pthr$ such that $\Lab(\pth{\pi'}) = \Lab(\pth{\rho'})$ and
   for every path $\pth{\rho''}$ such that
   $\pthr\suff\pth{\rho''}\psuff\pth{\rho'}$ there exists a path
   $\pth{\pi''}$ with $\pthp\suff\pth{\pi''}\psuff\pth{\pi'}$
   such that $\Lab(\pi'')=\Lab(\rho'')$.
   By induction, this implies $\pth{\rho'}\sat\pfrm{\psi''}$ and
   $\pth{\rho''}\sat\pfrm{\psi'}$ for all
     $\pthr\suff\pth{\rho''}\psuff\pth{\rho'}$.
   Hence $\pthr\models\pfrmp$.

``If'':
   Suppose that $s \not\ls t$.
   Then, without loss of generality, there exists a maximal
   path $\pthr$ from $t$ such that for all maximal paths $\pthp$ from
   $s$ it holds that $\Lab(\pthp)\not=\Lab(\pthr)$; we define an
   $\LTLminX$ formula $\pfrmp$ such that $s\sat\pfrmp$, while
   $t\not\sat\pfrmp$.

   First, we define for every colour $C$, which is a subset of
   $\Props$, a formula $\psi(C)$ with the property that $\pi \sat
   \psi(C)$ iff the first state of $\pi$ has colour $C$. (A possible
   definition of $\psi(C)$ would be $\bigwedge_{p\in C} p \wedge
   \bigwedge_{p \not\in C} \neg p$; however, one can economise on the
   cardinality of this conjunction by including only one conjunct for
   every other colour $D$ that actually occurs in the underlying
   Kripke structure---this way we meet the cardinality restriction
   imposed in Section~\ref{CTL}.)
   For every maximal path $\pthp$ from $s$ such that $\Lab(\pthr)$ is
   not a prefix of $\Lab(\pthp)$, let
   \begin{equation*}
     \pfrmp_{\pthp}=
       (\cdots((\psi(C_0))\Until (\psi(C_1)))\Until \cdots )\Until (\psi(C_k))
   \enskip,
   \end{equation*}
   where $C_0,C_1,\dots,C_k$ is the shortest prefix of
   $\Lab(\pthr)$ that is not also a prefix of $\Lab(\pthp)$.
   For every maximal path $\pthp$ from $t$ such that $\Lab(\pthr)$ \emph{is}
   a prefix of $\Lab(\pthp)$, let
   \begin{equation*}
     \pfrmp_{\pthp}= \compl
       (\cdots((\psi(D_0))\Until (\psi(D_1)))\Until \cdots )\Until (\psi(D_k))
   \enskip,
   \end{equation*}
   where $D_0,D_1,\dots,D_k$ is the shortest prefix of
   $\Lab(\pthp)$ that is not also a prefix of $\Lab(\pthr)$.
   Note that in either case we have
     $\pthr\sat\pfrmp_{\pthp}$
   while
     $\pthp\not\sat\pfrmp_{\pthp}$.
   Now, define $\pfrmp$ by
   \begin{equation*}
     \pfrmp =
       \compl\Meet\{
         \pfrmp_{\pthp} \mid \text{$\pthp$ a maximal path from $s$}
       \}
   \enskip.
   \end{equation*}
   It is not hard to check that in a Kripke structure with less then
   $\kappa$ states, for $\kappa$ an infinite cardinal, less than
   $\kappa$ of the formulas $\psi_\pi$ are different.
   Now, since $\pthr$ is a path from $t$ such that $\pthr\not\sat\pfrmp$,
   it follows that $t\not\sat\pfrmp$.
   On the other hand, since $\pthp\not\sat\pfrmp_{\pthp}$, it follows
   that $\pthp\sat\pfrmp$ for all paths $\pthp$ from $s$, and hence
   $s\sat\pfrmp$.
\end{proof}

\noindent
In order to lift this notion of equivalence from Kripke structures to
LTSs, consider a \emph{trivial colouring} \mbox{\fsc T}, assigning the
same colour to all states in an LTS, and write $s =_T^\lambda t$
if $s$ and $t$ have the same complete \mbox{\fsc T}-coloured traces.
In \cite{Gla93}, $=_T^\lambda$ was called
\emph{divergence sensitive trace equivalence}.
The following counterpart of Theorem~\ref{agreement} indicates that
$=_T^\lambda$ is on LTSs what $\ls$ is on Kripke structures:

\begin{thm}\label{trace agreement}
On a consistent L$^2\!$TS $\ls$ equals $=_T^\lambda$.
\end{thm}

\begin{proof}
If $\pi$ is a path from a state $s$ and $\rho$ a path from $t$
in a consistent {\dlts} $(S,\Lab,\rightarrow)$, then
$$\Lab(\pi) = \Lab(\rho) \Leftrightarrow \Lab(s) = \Lab(t) \wedge
 \mbox{\fsc T}(\pi) = \mbox{\fsc T}(\rho)$$
where $\Lab(\pi)$ denotes the $\Lab$-coloured trace in the associated
Kripke structure (thus, forgetting the actions) and $\mbox{\fsc T}(\pi)$
denotes the trivially coloured trace in the associated LTS (thus,
keeping the visible actions, but forgetting the colours).
This is an immediate consequence of the definition of consistency, and
it immediately implies the theorem.
\end{proof}

\noindent
In order to make LTS-to-\dlts{} transformations useful for applying
LTL on LTSs they should be required to preserve and reflect
$=_T^\lambda$---the transformation of \cite{DV95} trivially has this
property. We then obtain:

\begin{cor} Let $s$ and $t$ be states in an LTS, and let
  $\eta$ be an LTS-to-L$^2\!$TS transformation preserving and
  reflecting $=_T^\lambda$. Then
    $s =_T^\lambda t$
  iff
    $s\models^\eta \pfrmp\Leftrightarrow t\models^\eta\pfrmp$
  for all \LTLminX{} formulas $\pfrmp$.
\end{cor}

\noindent
The very same counterexample as used in Section~\ref{sec:pcmp} shows
that $=_T^\lambda$ fails to be a congruence for $\|$:
we have $0 =_T^\lambda \Delta 0$, yet
$0 \| a \not=_T^\lambda \Delta 0 \|a$. We proceed to
characterise the coarsest congruence for $\|$ that is included in
$=_T^\lambda$. We write $s =_T^{\Delta\lambda} t$ if $s$ and $t$ have
the same complete \mbox{\fsc T}-coloured traces as well as the same
divergent \mbox{\fsc T}-coloured traces; by analogy with the branching
bisimulation variants we propose to call $=_T^{\Delta\lambda}$
\emph{trace equivalence with explicit divergence}.

\begin{thm}
$=_T^{\Delta\lambda}$ is the coarsest congruence for
$\pcmpsym$ that is included in $=_T^\lambda$.
\end{thm}

\begin{proof}
Let $T(s)$ denote the set of \mbox{\fsc T}-coloured traces of a state
$s$, $T^\lambda(s)$ its set of complete \mbox{\fsc T}-coloured traces,
and $T^\Delta(s)$ its set of divergent ones.  Clearly $T^\Delta(s)
\subseteq T^\lambda(s) \subseteq T(s)$.  Note that $T(s)$ is
completely determined by $T^\lambda(s)$, namely as its set of initial
prefixes.  Furthermore, let $T^*(s)$ denote the set of finite
\mbox{\fsc T}-coloured traces of $s$ and $T^\infty(s)$ its set of
infinite ones.  Also $T^*\hspace{-1pt}(s)$ and $T^\infty\hspace{-1pt}(s)$ are
completely determined by $T^\lambda\hspace{-1pt}(s)$, and
$T^\infty\hspace{-1pt}(s)\mathbin\subseteq T^\lambda\hspace{-1pt}(s)$. 
For any two sets of sequences $S$ and $T$, let $S \| T$ denote the set
of those sequences which can be obtained by interleaving a sequence of
$S$ with a sequence of $T$. Now we have
$$\begin{array}{r@{~=~}l}
T(s\|t) & T(s) \| T(t) \\
T^*(s\|t) & T^*(s) \| T^*(t) \\
T^\infty(s\|t) & T^\infty(s) \| T(t) \cup T(s) \| T^\infty(s)\\
T^\Delta(s\|t) & T^\Delta(s) \| T^*(t) \cup T^*(s) \| T^\Delta(s)\\
T^\lambda(s\|t) & T^\infty(s\|t) \cup T^\Delta(s \| t) \cup
T^\lambda(s) \| T^\lambda(t).
\end{array}$$
This implies that $=_T^{\Delta\lambda}$ is a congruence.
By construction it is included in $=_T^{\lambda}$.

Now let $\sim$ be any congruence for $\|$ that is included in
$=_T^{\lambda}$, and assume $s \sim u$. We need to show that $s
=_T^{\Delta\lambda} u$. We know already that $T^\lambda(s) =
T^{\lambda}(u)$. So let $\sigma \in T^\Delta(u)$. By symmetry, it
suffices to show that $\sigma \in T^\Delta(s)$.  Let $a$ be an action
that does not occur in any path from $s$.  Since $\sim$ is a
congruence for $\pcmpsym$, we have $s\pcmp a \sim t\pcmp a$, where $a$
is the state from Example~\ref{deadlock}.  As $\sim$ is included in
$=_T^{\lambda}$ we obtain $s\pcmp a =_T^{\lambda} t\pcmp a$. Since
$\sigma \in T^\Delta(u)$ and the empty trace $\varepsilon$ is in
$T^*(a)$, we have $\sigma \in T^\Delta(u\|a) \subseteq T^\lambda(u\|a)
= T^\lambda(s\|a)$.  Since $\varepsilon \not\in T^\lambda(a)$ it must
be that $\sigma \in T^{\Delta}(s\|a)$ and hence $\sigma\in T^\Delta(s)$.
\end{proof}

\noindent
So far the situation is analogous with the branching time
case. However, from here on the development is different.  Adding the
$\infty$-modality to {\LTLminX} does not merely add the expressiveness
to the logic to make it characterise $=_T^{\Delta\lambda}$.
Instead {\LTLD} (obtained from \LTLminX{} by adding the
$\infty$-modality) characterises a strictly finer equivalence. We
define \emph{\Lab-coloured deadlock traces} as \Lab-coloured traces
that stem from finite maximal paths, i.e.\ paths ending in a deadlock
state, and for $s,t$ states in a Kripke structure
$(S,\Lab,\rightarrow)$ we write $s \approx_{\lab}^{\Delta\delta} t$ if
$s$ and $t$ have the same complete \Lab-coloured traces, the same
divergent \Lab-coloured traces, and the same \Lab-coloured deadlock traces.
Likewise, for $s,t$ states in an LTS we write $s =_T^{\Delta\delta} t$ if
$s$ and $t$ have the same complete \mbox{\fsc T}-coloured traces, the
same divergent \mbox{\fsc T}-coloured traces, and the same \mbox{\fsc
T}-coloured deadlock traces. In \cite{Gla93}, $=_T^{\Delta\delta}$ was
called \emph{divergence sensitive completed trace equivalence}.
In light of the proof of Theorem~\ref{trace agreement} it is
straightforward to establish that on a consistent \dlts{} the
preorders $\approx_{\lab}^{\Delta\delta}$ and $=_T^{\Delta\delta}$
coincide.

\begin{thm}
$s \approx_{\lab}^{\Delta\delta} t$ iff
$s\models \pfrmp \Leftrightarrow t\models\pfrmp$
for all \LTLD{} formulas $\pfrmp$.
\end{thm}

\begin{proof}
Let $\Lab^{\;\delta}(\pi)$ be the \Lab-coloured trace of a path $\pi$ as
given in Definition~\ref{stuttering}, but with a symbol $\delta$
tagged at the end iff $\pi$ is finite and maximal (i.e.\ ending in deadlock).
Then $s \approx_{\lab}^{\Delta\delta} t$ iff for every path $\pi$ from $s$
there is a path $\rho$ from $t$ such that $\Lab^{\;\delta}(\pi) =
\Lab^{\;\delta}(\rho)$, and vice versa.

``Only if'':
To show that
     $s \approx_{\lab}^{\Delta\delta} t$ implies
        $s\models \pfrmp \Leftrightarrow t\models\pfrmp$,
   it suffices to prove that
     if $\Lab^{\;\delta}(\pthp)=\Lab^{\;\delta}(\pthr)$
     then $\pthp\models \pfrmp \Leftrightarrow \pthr\models\pfrmp$.
This proceeds exactly as in the proof of Theorem~\ref{LTLchar}, except
that there is one extra case to consider, namely that $\pfrmp = \infty$:
Suppose $\pi \models \infty$. Then $\Lab^{\;\delta}(\pi)$ does not end
in $\delta$, so $\Lab^{\;\delta}(\rho)$ does not end in $\delta$, so
$\rho \models \infty$.

``If'':
   Suppose that $s \not\approx_{\lab}^{\Delta\delta} t$.
   Then, without loss of generality, there exists a maximal
   path $\pthr$ from $t$ such that for all maximal paths $\pthp$ from
   $s$ it holds that $\Lab^{\;\delta}(\pthp)\not=\Lab^{\;\delta}(\pthr)$.
   As in the proof of Theorem~\ref{LTLchar} we define an $\LTLminX$
   formula $\pfrmp$ such that $s\sat\pfrmp$, while $t\not\sat\pfrmp$.
For $\pi$ a maximal path from $s$ such that $\Lab(\pi) \neq \Lab(\rho)$,
we define the formula $\pfrmp_\pi$ exactly as in the proof of 
Theorem~\ref{LTLchar}. In case $\Lab(\pi)= \Lab(\rho)$ but
$\Lab^{\;\delta}(\pthp)\neq\Lab^{\;\delta}(\pthr)$ we take $\pfrmp_\pi$
to be $\infty$ or $\neg\infty$. The definition of $\pfrmp$ remains the same.
\end{proof}

\begin{cor}
  Let $s$ and $t$ be states in an LTS, and let $\eta$ be an
  LTS-to-L$^2\!$TS transformation preserving and reflecting
  $=_T^{\Delta\delta}\!$. Then $s =_T^{\Delta\delta} t$ iff
  $s\models^\eta \pfrmp\Leftrightarrow t\models^\eta\pfrmp$ for all
  \LTLD{} formulas $\pfrmp$.
\end{cor}

\noindent
The deadlock extension of Definition~\ref{def:deadlockextension} gives
the same result.

\begin{thm}
  Let $s$ and $t$ be states in an LTS, and let $\eta$ be an
  LTS-to-L$^2\!$TS transformation preserving and reflecting
  $=_T^{\Delta\delta}$. Then
    $s =_T^{\Delta\delta} t$
  iff
    $s\models^{\de\circ\eta} \pfrmp
       \Leftrightarrow
     t\models^{\de\circ\eta}\pfrmp$
  for all \LTLminX{} formulas $\pfrmp$.
\end{thm}

\begin{proof}
Just like Corollary~\ref{modal char}, this follows immediately from
the observations that $s \approx_{\lab}^{\Delta\delta} t$ within a
Kripke structure $\KS$ iff $s \approx_{\lab}^{\Delta\delta} t$ within
the Kripke structure $\de(\KS)$ (cf.\ Theorem~\ref{deadlock extension}),
and that on total Kripke structures the equivalence relations
$\approx_{\lab}^{\Delta\delta}$ and $\ls$ coincide.
\end{proof}

Kaivola \& Valmari \cite{KV92} study equivalences on LTSs with the
property that under all plausible transformations of LTSs into Kripke
structures two equivalent states (transformed into states of Kripke
structures) satisfy the same formulas in either {\LTLminX}  or
{\LTLD}. They characterise the coarsest such congruences for a
selection of standard process algebra operators---including the
merge, but also a partially synchronous parallel composition as well
as nondeterministic choice---as \emph{NDFD}-equivalence
(for {\LTLminX}) and \emph{CFFD}-equivalence (for {\LTLD}).
In turns out that neither $=_T^{\Delta\lambda}$ nor
$=_T^{\Delta\delta}$ are congruences for the partially synchronous
parallel composition, or for nondeterministic choice.
Hence to satisfy the requirement of being a congruence for these
operators, \emph{NDFD}-equivalence is necessarily finer than
$=_T^{\Delta\lambda}$, and \emph{CFFD}-equivalence is necessarily
finer than $=_T^{\Delta\delta}$.
The question of raising the expressiveness of {\LTLminX} to the level
where it characterises NDFD- or CFFD-equivalence directly remains open.

\section{Conclusion}

\noindent
  In this paper we enabled $\CTLminX$ and $\CTLSminX$ to be used as
  logics on labelled transition systems (LTSs) while taking deadlock
  behaviour into account. This could be accomplished by adding a
  modality to $\CTLSminX$, by adapting the semantics of the
  $\mathsf{G}$-modality (in $\CTLminX$), or by adapting the
  translations from \cite{DV95} from LTSs to Kripke structures. We
  have shown that these approaches all lead to equally expressive
  logics on LTSs.  Our work allows the rich tradition of verification
  by equivalence checking to be combined with the full expressive
  power of $\CTLSminX$. Taking advantage of this possibility is left
  for further research.

\subsubsection*{Acknowledgements}
We are grateful to the referees for their many helpful suggestions.

\bibliographystyle{plain}

\end{document}